\documentclass[aps,onecolumn,floatfix,superscriptaddress,showkeys,
showpacs,amsfonts,amssymb,amsmath]{revtex4}
\usepackage{amsmath}
\usepackage{graphicx}
\usepackage{amssymb}
\usepackage{amsfonts}
\usepackage{bm}
\usepackage{latexsym}

\usepackage[latin1]{inputenc}

\def\beq{\begin{eqnarray}}
\def\eeq{\end{eqnarray}}
\def\ln{\,\mbox{ln}\,}

\def\al{\alpha}
\def\be{\beta}

\def\ga{\gamma}
\def\de{\delta}
\def\vp{\varepsilon}
\def\ep{\epsilon}

\def\La{\Lambda}

\def\Om{\Omega}

\begin{document}

\vskip 6mm

\title{Testing dark matter warmness and quantity via the reduced 
relativistic gas model}
\author{Julio\ C.\ Fabris}
\affiliation{\ Departamento de F{\'\i}sica -- CCE, 
Universidade Federal do Esp{\'\i}rito Santo 
\\
Vit\'oria, CEP: 29060-900, ES, Brazil}
\author{Ilya\ L.\ Shapiro}
\affiliation{\ Departamento de F\'{\i}sica -- ICE,
Universidade Federal de Juiz de Fora
\\
Juiz de Fora, CEP: 36036-330, MG,  Brazil}
\author{A.\ M.\ Velasquez-Toribio}
\affiliation{\ Departamento de F{\'\i}sica -- CCE, Universidade Federal
do Esp{\'\i}rito Santo 
\\
Vit\'oria, CEP: 29060-900, ES, Brazil}
\date{\today}

\begin{abstract}
We use the framework of a recently proposed model of reduced 
relativistic gas (RRG) to obtain the bounds for $\Omega$'s 
of Dark Matter and Dark Energy (in the present case, a 
cosmological constant), taking into consideration an arbitrary 
warmness of Dark Matter. An equivalent equation of state has
been used by Sakharov to predict the oscillations in the matter 
power spectrum. Two kind of tests are accounted for in what 
follows, namely the ones coming from the dynamics of the 
conformal factor of the homogeneous and isotropic 
metric and also the ones based on linear cosmic perturbations. 
The RRG model demonstrated its high effectiveness, permitting 
to explore a large volume in the space of mentioned parameters 
in a rather economic way. Taking together the results of such
tests as Supernova type Ia (Union2 sample), $H(z)$, CMB 
($R$ factor), BAO and LSS (2dfGRS data), we confirm that 
$\La$CDM is the most favored model. At the same time, for the 
2dfGRS data alone we found that an alternative model with
a very small quantity of a Dark Matter is also viable. This
output is potentially relevant in view of the fact that the LSS 
is the only test which can not be affected by the possible 
quantum contributions to the low-energy gravitational action. 
\end{abstract}


\maketitle
\vskip 6mm

\section{\bf  Introduction}                           %

According to the standard estimates of the energy balance of 
the present-day Universe, the relative energy densities of the 
Dark Energy and Dark Matter (DM) are close to $\Om_\La^0=0.7$ 
and $\Om_{DM}^0=0.25$, respectively, while the visible (more 
precisely, baryonic) matter is represented by a modest less 
than 5\% of the total energy density \cite{riess,hannestad}.
Both Dark Energy and DM represent somehow mysterious 
components, but for different reasons. The main candidate 
to be Dark Energy is a cosmological constant (CC), however the 
cosmological constant problem is considered as one of the most 
difficult problems of the modern fundamental physics. The main 
CC problem is due to the conflict between the overall value of 
CC which is likely causing an acceleration of the present-day 
universe, and several much greater contributions to it which 
can be evaluated at the particle physics scale (see 
\cite{weinberg89} for a review). At the moment, no solution 
of this problem is known. On the other hand, there is another 
difficulty, which is known as a problem of coincidence. The 
problem is to explain why the universe started to accelerate 
only recently. This problem can 
be solved or at least alleviated in the two distinct ways. 
First, one can replace the CC term by some artificially 
designed substance (e.g. quintessence) which can adjust 
its energy density according to the expansion of the universe. 
Indeed, this approach is essentially worsening the situation 
with the main CC problem, because the amount of the requested 
fine-tuning becomes even greater and goes beyond the original 
one, which is due to the hierarchy between the particle physics 
scale and cosmic scale \cite{nova}. The second approach is 
much less explored, despite it looks much simpler. It assumes 
changing the energy balance of the universe such that in the 
new model $\Om_\La^0$ becomes closer to unity. In this case 
the acceleration starts earlier and the coincidence problem 
becomes partially resolved. 

The unique way to increase  $\Om_\La^0$ is by the expense 
of $\Om_{DM}^0$, and from the first sight this idea does not 
look realistic. The well-known reason is that the DM is required 
by a well-established set of observational data (see, e.g.,  
books \cite{KT,Dod} and reviews 
\cite{DM1,DM2,susyDM,Roos}). At the same time it is 
interesting to verify the mentioned possibility without 
prejudice, especially taking the possible warmness of the 
DM into account \cite{first,dodelson}. 
The exploration of the wide set of observational data is 
a complicated task. In particular, one has to choose such a 
model which could take care about possible DM warmness 
and, at the same time, keep the amount of requested 
calculations under control. It is also desirable to be able 
to implement some new physical input into this 
model, for the case we will like to make it at the next 
stages of investigation. All in all, we need sufficiently 
simple, but yet physically reasonable model for a warm DM 
(WDM). Fortunately, a very useful simplified model of this 
sort, called reduced relativistic gas (RRG), has been 
suggested recently in \cite{FlaFlu,sWIMPs}. An important 
historical note has to be done. After the first version of 
this paper has been submitted, we have learned from 
\cite{Grishchuk} that the equivalent simplified model, 
interpolating between radiation and matter epochs, has been 
used by A.D. Sakharov in the famous work \cite{Sakharov}, 
where the oscillations in the matter power spectrum were 
discovered for the first time. 

The idea of RRG 
is to treat WDM as an approximately Maxwell-distributed 
ideal gas of massive particles. As far as the large-scale 
description of the Universe concerns mainly the equation of 
state of the DM, it is not supposed to be sensitive to the 
non-collisional nature of the DM particles and to their 
anisotropic distribution at the astrophysical scale. 
Hence, there can not be much difference between the RRG 
description from one side and the one based on relativistic 
Fermi distribution from another one. The same is expected 
to be true, up to some extent, for the more complete approach 
based on the Boltzmann-Einstein system of equations \cite{Dod}. 

The expression ``approximately Maxwell'' concerning the 
RRG model means we take an approximation where all DM 
particles have equal kinetic energies. As it was shown 
in \cite{FlaFlu}, qualitatively and numerically the equation 
of state suffers only a very small modification from this 
approximation. At the same time, the model based on RRG 
admits simple analytical form for the equation of state 
and even enables one to analytically solve the Friedmann 
equation. The analysis of cosmic perturbations in the 
framework of RRG model and LSS data provided \cite{sWIMPs} 
an upper bound on the DM warmness which is very close to 
the one obtained from much more complicated analysis based 
on the standard Boltzmann-Einstein system 
\cite{szalay,ostriker1,ostriker2,ostriker3}. Thus, the 
RRG model represents a really useful tool for exploring 
the new cosmological models with the WDM. At the same 
time, it looks interesting to check this model on such 
sets of observational data as CMB, BAO and others. 

There is also a special motivation for us to explore the 
possibility of an alternative values of $\Om_{DM}$ with the WDM, 
which is related to the potentially relevant quantum effects of 
vacuum. The backgrounds of the cosmological and astrophysical 
applications of the theory with renormalization-group based 
quantum corrections were established in \cite{Gruni,CCG}. 
In the recent paper \cite{LRL} it was shown, from a general 
considerations based on covariance and dimensional arguments, 
that the form of these quantum corrections can be defined up to 
a single free parameter $\nu$. The recent analysis of the 
galaxies rotation curves in \cite{RotCurves} in the theory 
without DM, but with the mentioned quantum corrections to the 
Newton law was surprisingly successful. 
It was shown that such quantum corrections can compete with CDM 
in explanation of the rotation curves. The small amount of 
sufficiently warm DM does not affect these results and, 
therefore, we can think about weakening the standard 
restrictions to the DM in the theory with quantum corrections. 
On the other side, the cosmological model based on the same 
renormalization-group quantum corrections has been considered 
in [24,25]. The analysis of cosmic perturbations in 
this model [25] (see also 
\cite{CCwave}) demonstrated that the quantum contributions
almost do not affect the power spectrum, such that their 
impact on the LSS (2dfGRS) data is very weak. Therefore, 
if such quantum corrections do exist, the LSS data alone 
should manifest such possibility even in the zero-order 
approximation, that means without taking into account the 
quantum effects. At the same time, we should 
expect the quantum corrections to be relevant for other test, 
especially for lensing, Supernova, $H(z)$ and CMB.  In order 
to take this into account, one has to consider the 2dfGRS 
test separately. The positive result for the zero-order 
approximation is the case when the LSS part alone should be 
compatible with an alternative model, 
with smaller $\Om_{DM}$, while the full set of tests should 
uniformly indicate to the $\Lambda$CDM as unique possible option  
\textit{without} taking quantum effects into account. This is 
the only one possible output which can actually leave the 
chance for relevant effects of quantum corrections at the 
astrophysical and cosmological scale. 

We will present in this paper the RRG model of a WDM, 
characterized by the dark matter fractional density measured 
today $\Omega_{\rm dm0}$ and also by the parameter $b$, 
which specifies the degree of warmness of the DM component. 
The model will be confronted against background observational 
data (Supernova type Ia, Baryonic Acoustic Oscillation,
Cosmic Microwave background $R$ parameter and the age of the 
universe as function of $z$, $H(z)$) and the matter power 
spectrum at the linear regime. We will see that the background 
tests, in general, favor a scenario of the $\Lambda$CDM model, 
with $b \approx 0$, and $\Omega_{\rm dm0} \approx 0.25$, even if 
for a given specific tests some particularities may appear. 
However, the matter power spectrum alone favors a model with 
a very small value of $\Omega_{\rm dm0}$, giving the room for 
some degree of warmness. In view of the discussion presented 
above this result can be seen as a sign for a possibility of 
a new concordance scenario if the quantum corrections are 
taken into account.

The paper is organized as follows. In Sect. 2 we present 
a brief necessary review of the RRG model. The contents of 
this section is essentially the same as \cite{sWIMPs} and we 
include it only for making all considerations in the paper 
more consistent. In Sect. 3, we discuss linear perturbations 
of matter including baryon component. The observational
data which are used in our analysis are presented in Sect. 4. 
Finally, in Sect. 5 we draw our conclusions and discuss
further perspectives for exploring an alternative 
concordance model with quantum corrections.

\section{\large\bf  Background notions of the RRG model for WDM}

In this section we present the necessary elements of the 
RRG model \cite{FlaFlu,sWIMPs}. 

\subsection{Equation of state and zero-order cosmology}

Consider the cosmological 
model which is based on the relativistic gas of massive 
particles representing WDM, plus baryonic matter, radiation 
and cosmological constant. As a simplification, it is assumed 
that the relativistic gas is composed by particles with equal 
kinetic energy, or equal speed $\be=v/c$. By elementary means 
one can obtain the following equation of state, which links 
the pressure $P$ and energy density $\rho$ of such gas, 
\beq
P &=& \frac{\rho}{3}\,\Big[1-\Big(\frac{mc^{2}}{\vp}\Big)\Big]^{2}
\,=\, \frac{\rho}{3}\,\Big(1-\frac{\rho^2_{d}}{\rho^2}\Big)\,.
\label{2}
\eeq
In this formula $\vp$ is the kinetic energy of the individual 
particle, $\vp=mc^2/\sqrt{1-\be^2}$. Furthermore, 
$\rho_{d}=\rho_{d0}^{2}(1+z)^{3}$  is the mass (static 
energy) density. One can use one or another form of the 
equation of state (\ref{2}), depending on the situation. 
For example, in order to explore the evolution of the 
conformal factor $\,a=a(t)\,$ in this model, one can take the 
second expression in  (\ref{2}) and using the conservation 
law find that the relative energy density for the relativistic 
gas is given by the expression
\beq
\Omega_{dm}(a) &=& 
\frac{\Omega_{dmo}}{a^{3}\,
\sqrt{1+b^{2}}}\sqrt{1+\frac{b^{2}}{a^{2}}}\,,
\label{4}
\eeq
where $\Omega_{\rm dm0}$ is the present-day dark matter density, 
and $b$ is related to particles speed. Indeed, $b \approx 0$ 
means that the particles are nonrelativistic. Furthermore, for 
small velocities one can easily obtain the relation $b \approx \be$ 
\cite{sWIMPs}. Finally, by using the equations presented above and 
the ones for other components, the Hubble parameter can be presented
in the form
\beq
&& H^{2}(a,\Om_{dm0},\Om_{b0},b,\Om_{r0}) 
\,=\, H_{0}^{2}\,\Big[\frac{\Om_{dmo}}{a^{3}\sqrt{1+b^{2}}}
\sqrt{1+\frac{b^{2}}{a^{2}}}+\frac{\Omega_{b0}}{a^{3}}
+\frac{\Omega_{r0}}{a^4}+\Omega_{\Lambda 0})
\Big]\,,
\eeq
where we assume that 
$$
\Omega_{\Lambda 0} = 1-\Omega_{\rm dm0}-\Omega_{bo}-\Omega_{r0}
$$
and use the value 
$\Omega_{r0}h^{2}=2.42 \times 10^{-5}$. In this work we consider 
only the space-time manifolds with the flat space section.

If taken alone, the RRG model provides a natural interpolation 
between the radiation-dominated and matter-dominated universe 
regimes. Due to the adiabatic expansion, the relativistic gas 
is cooling down and the universe can look as filled by 
ultrarelativistic gas at the early stage of expansion and almost 
as a dust at the final stage of this expansion. One can find more 
details about the RRG model in \cite{FlaFlu} and \cite{sWIMPs}.

\begin{figure}[htb]
\begin{center}
\includegraphics[height= 4.5 cm,width=5.0cm]{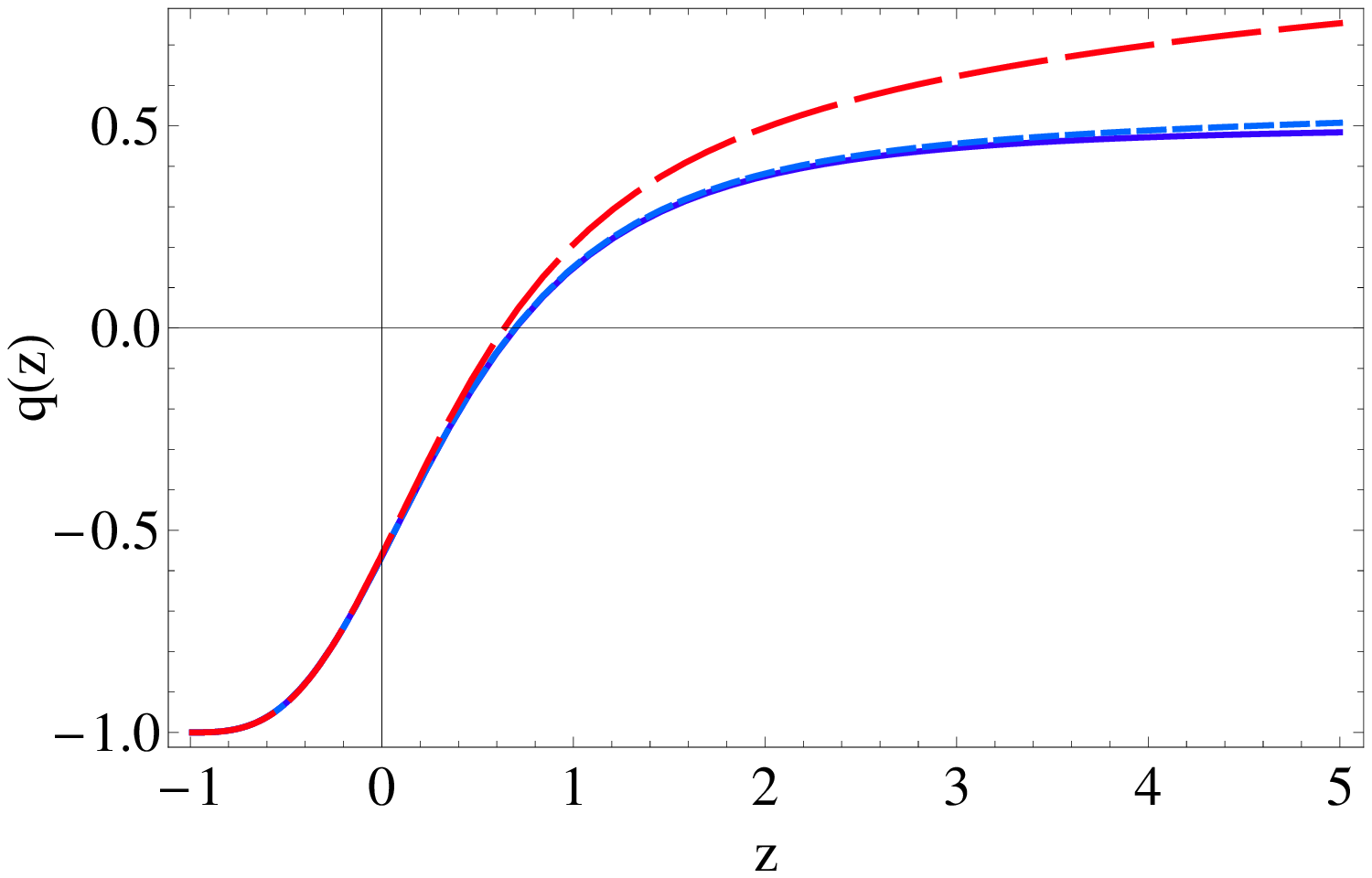}
\includegraphics[height= 4.5 cm,width=5.0cm]{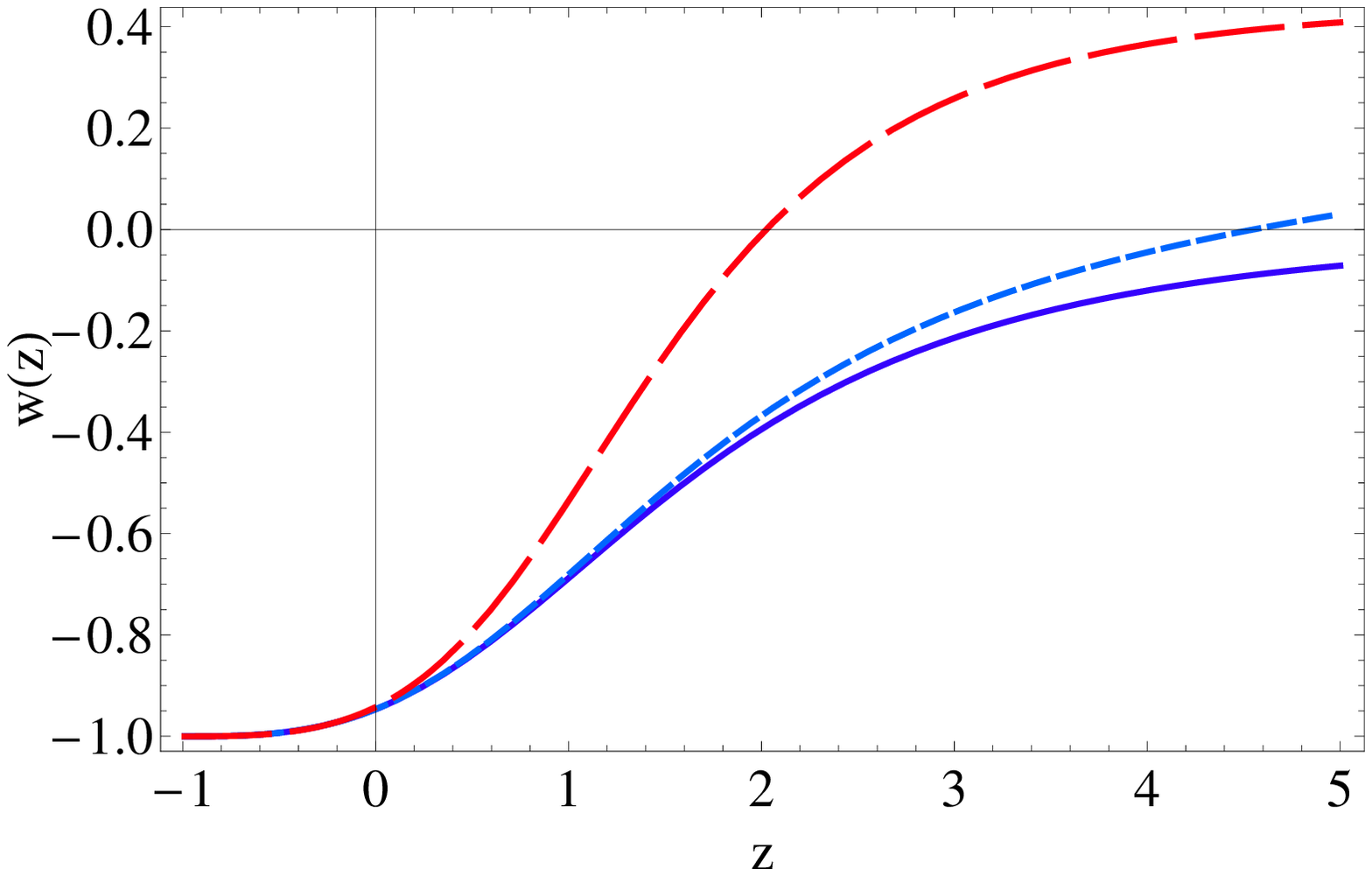}
\includegraphics[height= 4.5 cm,width=5.0cm]{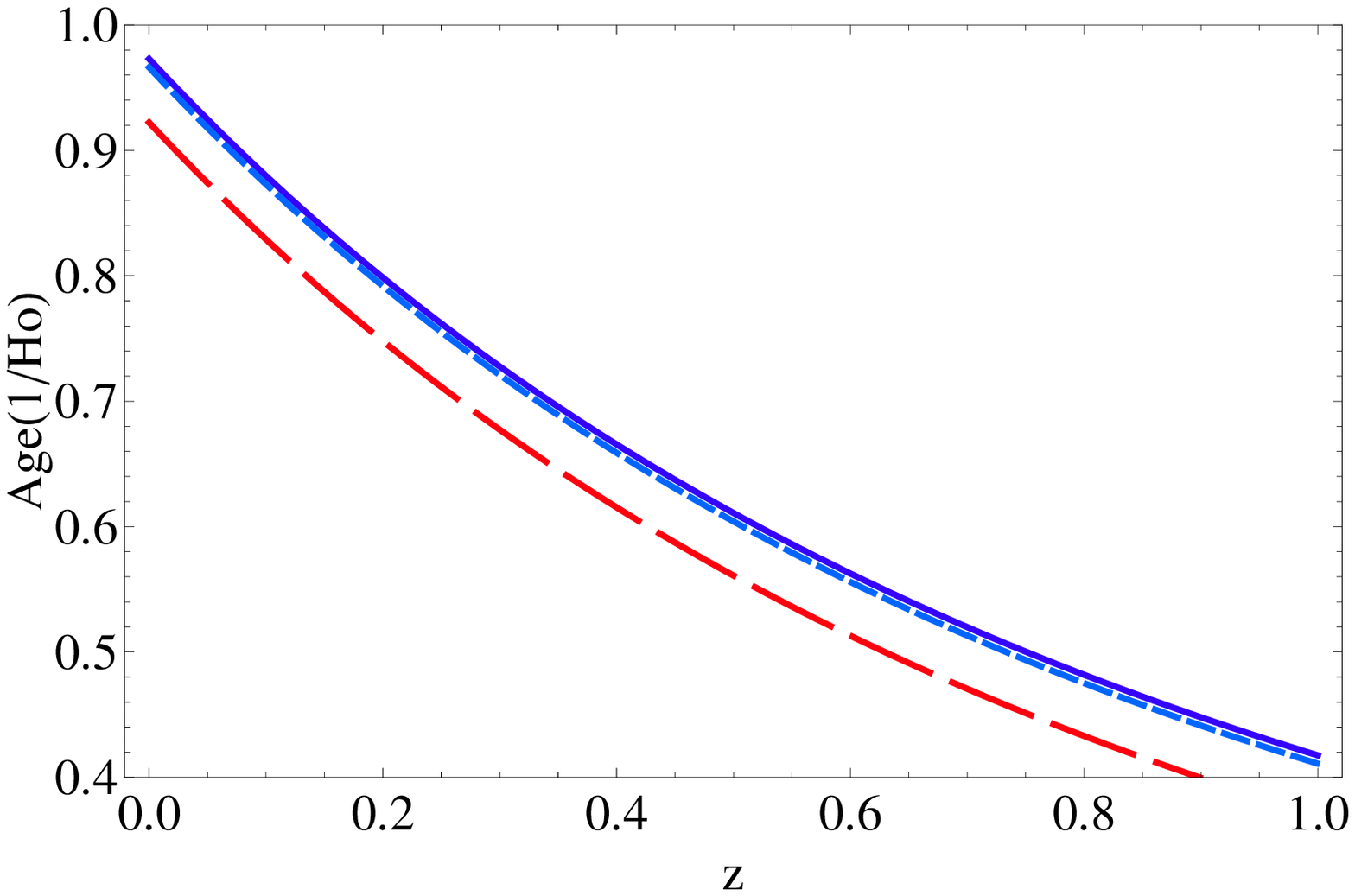}
\end{center}
\caption{The deceleration parameter (left), equation of state 
(center) and age universe (right) for different values of 
the parameter \ $b = 0, \,\, 0.01, \,\, 0.2$ \ 
(from top to bottom of the first plot). Here $w(z)$ is an 
accumulated parameter corresponding to the sum of the 
cosmological constant term and the matter content.  
In all cases here we used the values $\Omega_{b0}=0.04$ and 
$\Omega_{\rm dm0}=0.25$. 
The $\Lambda$CDM model corresponds to the value \ $b=0$. }
\label{fig1-1}
\end{figure}

In order to illustrate the effect of the warmness of the 
matter content of the universe, in Fig. \ref{fig1-1} we 
displayed the deceleration parameter $q(z)$, the equation of 
state $w(z)$ and the age universe in $H_{0}$ units. One can 
see that for $b \approx 0.1$ the function plot goes away from 
the $\Lambda$CDM plot with $(b = 0)$. The state function is 
the function more sensitive to the value of parameter $b$.

\section{\bf  Equations for Linear Perturbations}

Following the approach developed in our previous paper 
\cite{sWIMPs}, the dynamics of density perturbations has been 
analyzed in the synchronous gauge. The calculation of these 
perturbations is rather a standard routine, the only exception 
is the variation of the equation of state (\ref{2}), where one 
has to use the first form of this equation, taking 
\beq
\de P 
&=& \frac{\de \rho}{3}
\,\Big[1-\Big(\frac{Mac^{2}}{\vp}\Big)\Big]^{2}
\,=\, \frac{\de \rho}{3}\,\Big(1-\frac{\rho^2_{d}}{\rho^2}\Big)\,.
\label{2-vary}
\eeq
The reason is that in the framework of RRG model one has to 
provide kinetic energies of all particles to be equal and, 
therefore, we have no right to change the ratio $mc^{2}/\vp$. 
In other words, the variations of the energy density and of 
the rest energy density, $\de \rho_{d}$ and $\de \rho$, 
should be always proportional. 
By using this rule we arrive at the following system of 
equations for the linear perturbations {\bf\cite{sWIMPs}}:
\begin{eqnarray}
&& 
\delta_{b}'' 
\,+\, \Big(\frac{2}{a}+\frac{f'}{f}\Big)\delta_{b}'
\,=\, -\frac{3}{2}\frac{\Omega_{b}}{a^3 f^2}\delta_{b} - 
\frac{3}{2}\frac{\Omega_{dm}}{f^2}(1+3w)\delta_{dm}\,, 
\nonumber 
\\
&& \delta'_{dm} 
\,+\, (1+w)\Big(\frac{v}{f}-\delta'_{b}\Big) \,=\,0\,,
\\
&& (1+w)\,\Big[v'+(2-3w)\frac{v}{a}\Big] \,=\, 
\frac{k^2}{k_{0}^2}\,\frac{\omega\delta_{dm}}{a^{2}f}\,,
\nonumber
\label{eqns}
\end{eqnarray}
where $v$ is the 
peculiar velocity, $\delta_{b}$ is the baryonic matter 
density fluctuation and $\delta_{dm}$ is the relative 
dark matter density fluctuation. Other relevant 
definitions are
\beq
f &=& f(a) \,=\,
a \sqrt{\Omega_{dm}+\frac{\Omega_{b0}}{a^{3}}
+ 1-\Omega_{b0}-\Omega_{\rm dm0}}\,,
\nonumber
\\
w &=& w(a) = \frac{1-r(a)}{3}
\qquad  \mbox{and} \qquad
r(a)\,=\,\frac{1+b^2}{1+(b^2/a)^2}\,,
\label{pert}
\eeq
where $\Omega_{dm}$ is given by the Eq. (\ref{4}). 
The system above differs from the CDM model, because in our 
model the peculiar velocity and the state function $w(a)$
are non-zero. Additionally, the $CDM$ model interacts only 
through gravity and can be considered as a pressureless 
perfect fluid, unlike our model has a component of pressure 
given by (\ref{2-vary}). The solution of the system 
(5) was performed using a Mathematica-based code. 
To determine the power spectrum we use the BBKS transfer 
function \cite{bbks}. It is important to note that we also 
use the transfer function described in references 
\cite{bode,Riotto-Lyman} which are adapted for WDM particles, 
which depends on some specific inputs, obtaining the same 
results as when the BBKS transfer function is employed.

For the scale invariant spectrum, favored 
by the primordial inflationary scenario, the BBKS transfer 
function is given by 
\beq
T(k) &=& \frac{\ln(1+2.34 q)}{2.34 q}[1+3.89q+16.1q^{2}
+ 5.64q^{3}+6.71q^{4}]^{-1/4} 
\label{trans}
\\
\nonumber
\\
\mbox{where} && 
q(k) \,=\, \frac{k}{h\Gamma \,Mpc^{-1}}
\qquad \mbox{and}\qquad
\Gamma \,=\, \Omega_{m0}\,h \,
\exp\Big(-\Omega_{b0} - \frac{\Omega_{b0}}{\Omega_{m0}} \Big) \,.
\nonumber
\eeq

We have used the function \ $T(k)$ \ to impose the initial 
conditions of the system (5). 
The power spectrum is defined as usual, at $z=0$ we have 
\beq
P(k)\,=\, 
\mid \delta(k) \mid^{2} 
= AT(k)\, \Big[\frac{g(\Omega_{m0})}{\Omega_{T}}\Big]^{2}k\,,
\label{10}
\eeq
where $A$ is a normalization constant of the spectrum, 
which can be fixed from the spectrum of anisotropy of the 
cosmic microwave background radiation. The expression for 
$g(\Omega)$ is given by
\beq
g(\Omega) &=& \frac{5 \Omega}{
  2 (\Omega^{4/7} + 1.01(\frac{\Omega}{2} + 1) - 0.75)}
\eeq
The analysis of the linear density perturbations 
enables us to obtain the upper bound for 
the warmness of the DM in a very economic way \cite{sWIMPs}. 
In Fig. 1 we compare the 2dFGRS data with the power 
spectrum of our model for the usual $\La$CDM energy 
balance case. We used the best fit obtained using 
the $\chi^{2}$ in the next section. One can see that if
\ $b \approx 10^{-3}$, \  the predicted power spectrum is 
outside of the region of data shown in the plot. A detailed 
analysis of observational constraints is given in section 5.  

\begin{figure}[htp]
\begin{center}
\includegraphics[height= 9.0cm,width=9.5cm]{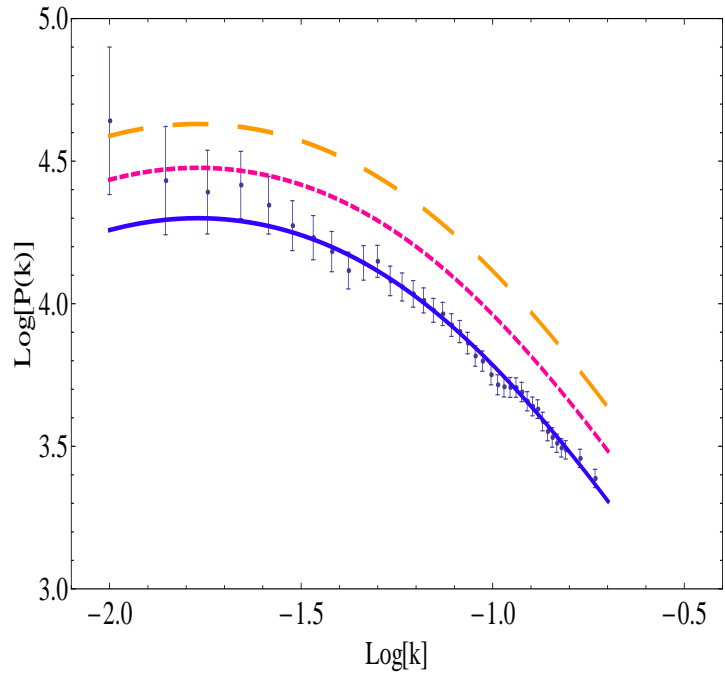}
\end{center}
\caption{
Power spectrum using 
$\Omega_{b0}=0.044$ and $\Omega_{\rm dm0}=0.24$, for three different 
values of \ $b=10^{-4},\,\,5.0 \times 10^{-3},\,\,10^{-2}$
\ (from bottom to top). It is easy to see that the last two values for 
$b$ provide the linear power spectrum, at small scale, which is 
incompatible to the 2dFGRS data.}
\end{figure}

\section{\bf  Constraints from Observational Data}

Nowadays there is a large number of observational tests focused 
on different aspects of the cosmological models. It is expected that 
the crossing of these different observational tests may lead to 
an expressive constraint on the free parameters of a given model, 
in some cases confirming the viability of the proposed scenario
or ruling it out. Here we consider essentially two kinds 
of tests: those focusing on the background functions, specially 
$H(z)$ and also the perturbative tests, which involve more 
profoundly the description of the matter/energy content of the 
model. Concerning the tests related to the background behavior 
only, we will consider the supernova type Ia data (Union2 sample), 
BAO, the position of the first acoustic peak in the CMB spectrum 
(the $R$ parameter) and the age of the old objects. In what 
concerns the perturbative tests, we will restrict ourselves 
to the matter power spectrum associated to linear perturbations.

The analysis of  statistics begins with the $\chi^2$ functions 
constructed following the general expression
\beq
\chi^2 = \sum_{i=n}^N\frac{(\rho_i^{th} - \rho_i^{ob})^2}{\sigma_i^2},
\eeq
where $N$ is the total number of observational data, $\rho_i^{th}$ 
are corresponding theoretical predictions and $\rho_i^{ob}$ 
represent the observational values, with an error bar given by 
$\sigma_i$. The probability distribution function is constructed 
from the $\chi^2$ function as
\beq
P(x^j) = A\,e^{-\chi^2(x^j)/2},
\eeq
where $x^j$ are the set of free parameters of the model and $A$ is a 
normalization factor. Within the RRG model, there are two free 
parameters, $\Omega_{\rm dm0}$ and $b$.  In what follows we describe 
the results for each of these tests.

\subsection{The CMB Shift Parameter}

We used the shift parameter, $R$, which relates the angular 
diameter distance to the last scattering surface with the angular 
scale of the first acoustic peak in the WMAP7 power spectrum, and 
is given by (for $k=0$) \cite{komatsu}
\beq
R(\Omega_{\rm dm0},\Omega_{b0},b) 
= \sqrt{\Omega_{m}}\int\limits_{0}^{1090}
\frac{dz}{E(\Omega_{\rm dm0},\Omega_{b0},b)} = 1.725\pm 0.018,
\eeq
where \
$E(\Omega_{\rm dm0},\Omega_{b0},b)
=H(\Omega_{\rm dm0},\Omega_{b0},b)/H_{0}$. \
It is important to point out that the measured value of $R$ is
model independent and, therefore, can be used as a test of 
cosmology. In general, the shift parameter alone is insufficient 
for this purpose.  
In ref. \cite{pn} it is shown how to derive the $R$ parameter  
in a model independent way. In order to include the CMB
shift parameter into the analysis, it is needed to integrate up to
the matter-radiation decoupling ($z \simeq 1090$), so that
radiation is no longer negligible and it should be properly taken 
into account. For our model we used the minimization of the 
$\chi^{2}$ statistic according to 
\begin{equation}
\chi^{2}_{R}
\,=\,\frac{[1.725-R(\Omega_{\rm dm0},\Omega_{b0},b)]^{2}}{0.018^{2}}\,.
\end{equation}
We used $\Omega_{b0}=0.04$. In the Fig. 7. one can see the results 
for our model. It is clear that this test alone can not severely 
restrict the space of parameters. In particular, this test does 
not allows us to determine an upper limit for the parameter $b$
for fixed $\Om$'s.

\subsection{BAO}

The primordial baryon-photon acoustic oscillations leave a
signature in the correlation function as
observed by Eisenstein et al. \cite{eisenstein}. This signature
provides us with a standard ruler which can be used to constrain
the following quantity
\begin{equation}
A(\Omega_{\rm dm0},\Omega_{b0},b) 
\,=\, \sqrt{\Omega_{m0}}\,
\big[E(z_1,\Omega_{\rm dm0},
\Omega_{b0},b)\big]^{-1/3}
\,\left[ \frac{1}{z_1} \int_0^{z_1} 
\frac{dz}{E(z,\Omega_{\rm dm0},\Omega_{b0},b)} \right]^{2/3},
\end{equation}
where  the observed value of $A$ 
is \ $A_{obs} = 0.469 \pm 0.017$ \ and \ $z_1 = 0.35$ \ is the 
redshift of the SDSS sample. The $\chi^{2}$ statistics may be 
computed from
\beq
\chi^{2}_{BAO}(\Omega_{\rm dm0},\Omega_{b0},b)
\,=\,\frac{[0.469-A(\Omega_{\rm dm0},0.044,b)]^{2}}{0.017^{2}}\,.
\eeq
Here we will use the \ $\Omega_{b0}=0.044$ \ value to build the 
regions confidence.
The results are presented in Fig.8. We can see that BAO does not 
impose an upper limit for the parameter b, but it restricts the value 
of $\Omega_{\rm dm0}$.

\subsection{H(z)}
The observational Hubble parameter depends on 
the redshift $z$ in the form
\beq
H(z) = - \frac{1}{1+z}\frac{dz}{dt}\,.
\eeq
\\
Then, $H(z)$ can be obtained as far as $dz/dt$ is known. 
Jimenez et al. \cite{ji} 
demonstrated the feasibility of the method while Simon 
et al. \cite{si} and Stern et al. \cite{stern} obtained $H(z)$ 
in the range of $0.09 < z < 1.8$. Another way to get values of 
$H(z)$ can be using the BAO peak position as a standard ruler 
in the radial direction. Using this prescription in the reference 
\cite{gazt} the authors determined $H(z)$ in $z=0.43$ and $z=0.24$.
We employ the eleven data of \cite{si,stern} and two data of 
\cite{gazt}. The best fit values for the model parameters from
observational Hubble data are determined by minimizing the quantity
\beq
\chi^{2}_{H}(\Omega_{\rm dm0},\Omega_{b0},b,H_{0})
=\sum_{i}^{13}{\frac{\big[H_{obs,i}
-H_{th}(\Omega_{\rm dm0},\Omega_{b0},b,H_{0})\big]^2}{\sigma_{i}^{2}}}\,.
\eeq
Here we use the value $\Omega_{b0}=0.04$. The free parameters are
$\Omega_{\rm dm0}$, $H_{0}$ and $b$. However, in our case, we have found 
that the one-dimensional probability distribution function in $H_{0}$ 
is quite narrow and the values $H_{0}$ are well approximated 
by the value $H_{0}=70.5$ (minimizing the $\chi^{2}_{H}$ considering 
the three free parameters).
Therefore, one has only two free parameters, namely $b$ and  
and the $\Omega_{\rm dm0}$ parameter. The results for this test are
presented in Fig. 9. In general, this test provides the results 
which are similar to the ones of BAO. In particular, this is 
because the parameter $b$ remains weakly constrained, but the
restrictions are strong for the value  of $\Omega_{\rm dm0}$.

\subsection{Compilation Union2}

The supernovae Ia data give us  the distance modulus 
\ $\mu$ \ to each supernova, that is given by
\beq
\mu \equiv m-M = 5 \log \Big[\frac{d_{L}}{Mpc}\Big]
\,+\,25\,,
\eeq
where $M$ is the absolute magnitude. The distance modulus can 
also be written as
\beq
\mu = 5\log_{10}D_{L}(z)+\mu_{0}\,,
\eeq
where $D_{L}=\frac{H_{0}d_{L}}{c}$ is the Hubble-free luminosity 
distance and $\mu_{0}$ is the zero point offset (which is an
 additional model-independent parameter) defined by 
\begin{equation}
\mu_0 \,=\, 5\log_{10}\Big(\frac{cH_0^{-1}}{Mpc}\Big) + 25 
\,=\, 42.38 \,-\, 5\log_{10}\, h\,.
\end{equation}

In the present paper we used the Union2 dataset including 557 data
of Amanullah et. al \cite{amanullah}, that includes 
the intermediate-z data observed during the first season of the 
Sloan Digital Sky Survey (SDSS)-II supernova survey \cite{kessler} 
and the high z data from the Union compilation \cite{kowalski}.

In our case the $\chi^{2}_{SN Ia}$ is given by 
\beq
\chi^{2}_{SN Ia}(p_{i}) = \sum_{i=1}^{n}{\frac{(\mu_{the}(p_{i},z_{i})
- \mu_{obs}(z_{i}))^{2}}{\sigma_{obs,i}^{2}}}\,.
\eeq

The $\chi^{2}$ function can be minimized with respect to the 
$\mu_{0}$ parameter, as it is independent
of the data points and the dataset. The $\sigma_{obs}$ are 
provided in \cite{amanullah} and include both the observational 
and intrinsic magnitude scatter. We assumed a Gaussian and 
uncorrelated probability distribution for the data. Expanding 
the equation above with respect to $\mu_{0}$, we obtain:
\beq
\chi^{2}(p_{i})_{SN Ia} 
= A(p_{i})-2\mu_{0}B(p_{i})+\mu_{0}^{2}C(p_{i})\,,
\eeq
which has a minimum for $\mu_{0} = B(p_{i})/C(p_{i})$, giving
\beq
\chi^{2}_{SN Ia,min} = \bar{\chi}^{2}_{SN Ia} = A(p_{i}) 
- \frac{B^{2}(p_{i})}{C(p_{i})}\,,
\eeq
where
\beq
A(p_{i})&=&\sum_{i}^{n}{\frac{(\mu_{th} - 
\mu_{obs}(p_{i},\mu_{0}=0))}{\sigma_{i}}^{2}}\,,
\\
B(p_{i}) &=& \sum_{i}^{n}{\frac{\mu_{th} - 
\mu_{obs}(p_{i},\mu_{0}=0)}{\sigma_{i}}}\,,
\\
C(p_{i})& =& \frac{1}{\sigma_{i}^{2}}\,.
\eeq
Let us note that the new $\bar{\chi}^{2}_{SN Ia}$ is independent 
on $\mu_{0}$ and can be minimized with respect to the parameters
of our theoretical model. The results are shown in the Fig. 10.
We can see that the parameters $b$ and $\Omega_{\rm dm0}$ are 
relatively constrained. In particular, the parameter $b$ has a 
maximum likelihood value for $b \approx 0.4$ (see Table 1).

\subsection{2dfGRS}

We used the 2dFGRS data \cite{cole} to compare the power spectrum of our
model and constrain the free parameters.  To do this, we minimize
$\chi^{2}$ defined as
\beq
\chi^{2}_{2dF}(\Omega_{\rm dm0}, \Omega_{b0},b)
=\sum_{i=1}^{39}\frac{[P((k_{i},\Omega_{\rm dm0},\Omega_{b0},b)
-P_{obs}(k_{i},\Omega_{\rm dm0},\Omega_{b0},b)]^{2}}{\sigma^{2}_{i}}\,.
\eeq
Here $P(k)$ is given by equation (\ref{10}) and the number of free 
parameters is two, since we fixed the value of $\Omega_{b0}=0.04$.
In Fig. 3 we present the PDF for our model. 
The impact that has the model on linear power spectrum 
is dominated by the $b$ parameter that is associated with the 
speed of dark matter particles. In the left figure we can see that
if $\Omega_{\rm dm0}$ is close to $\Lambda$CDM value, then, the $b$ value
is close to zero. But, if the value of $b$ is, for example, $10^{-4}$, 
then the dark matter fraction decreases significantly. The corresponding 
data can be seen in Table 1. 

\begin{figure}[!h]
\begin{center}
\includegraphics[height= 6.5cm,width=6.5cm]{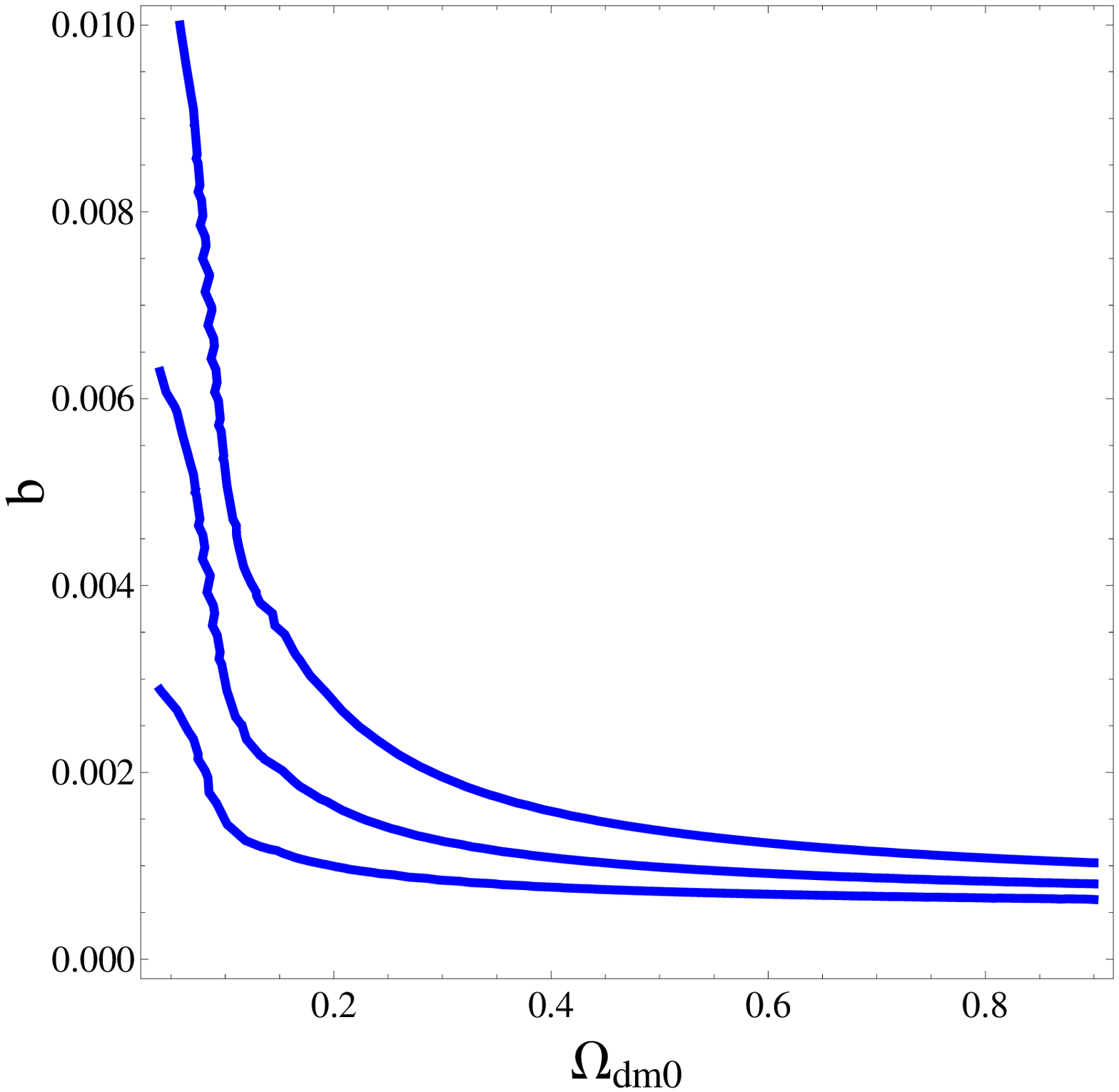}
\includegraphics[height= 6.5cm,width=6.5cm]{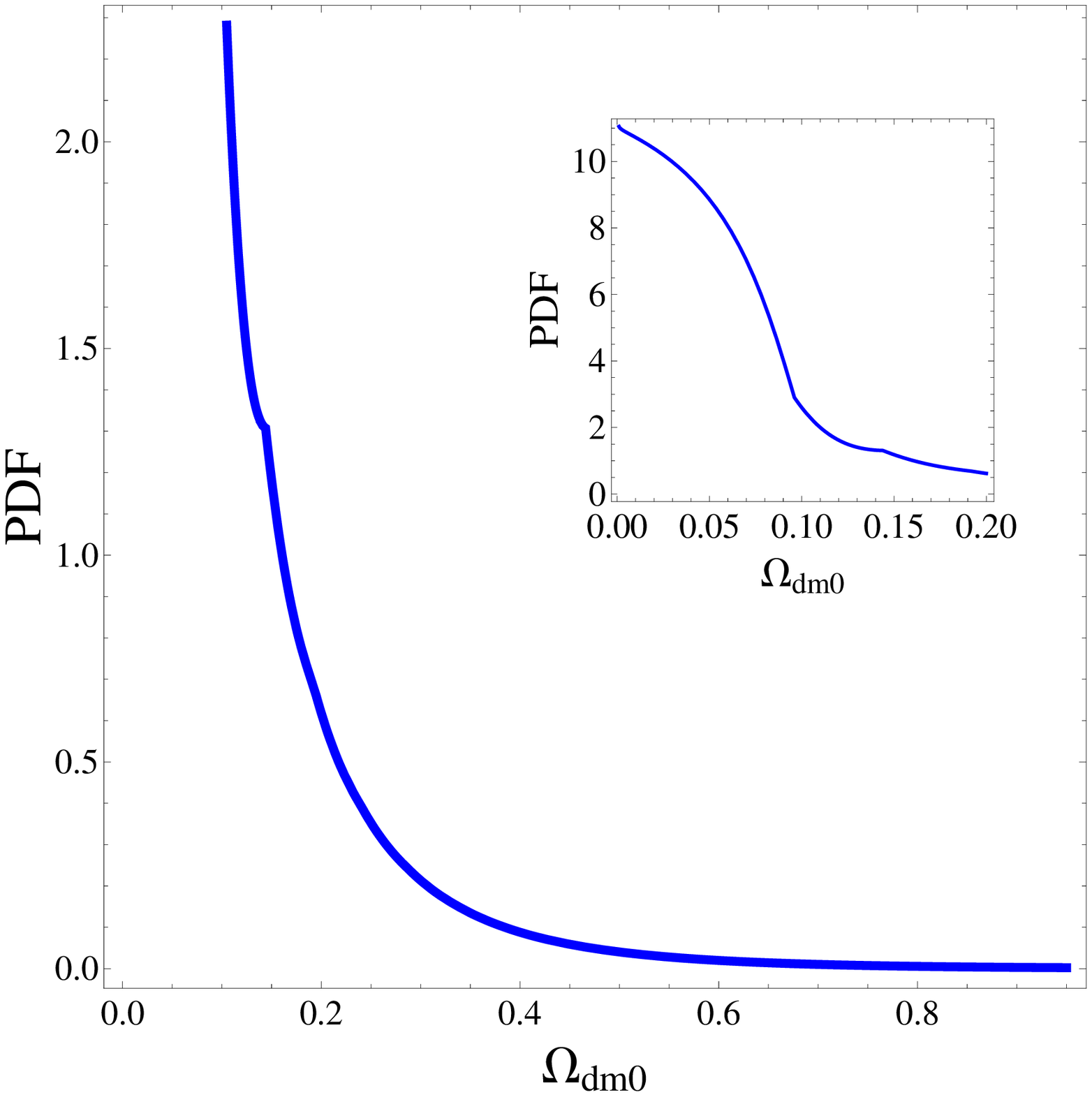}
\includegraphics[height= 6.5cm,width=6.5cm]{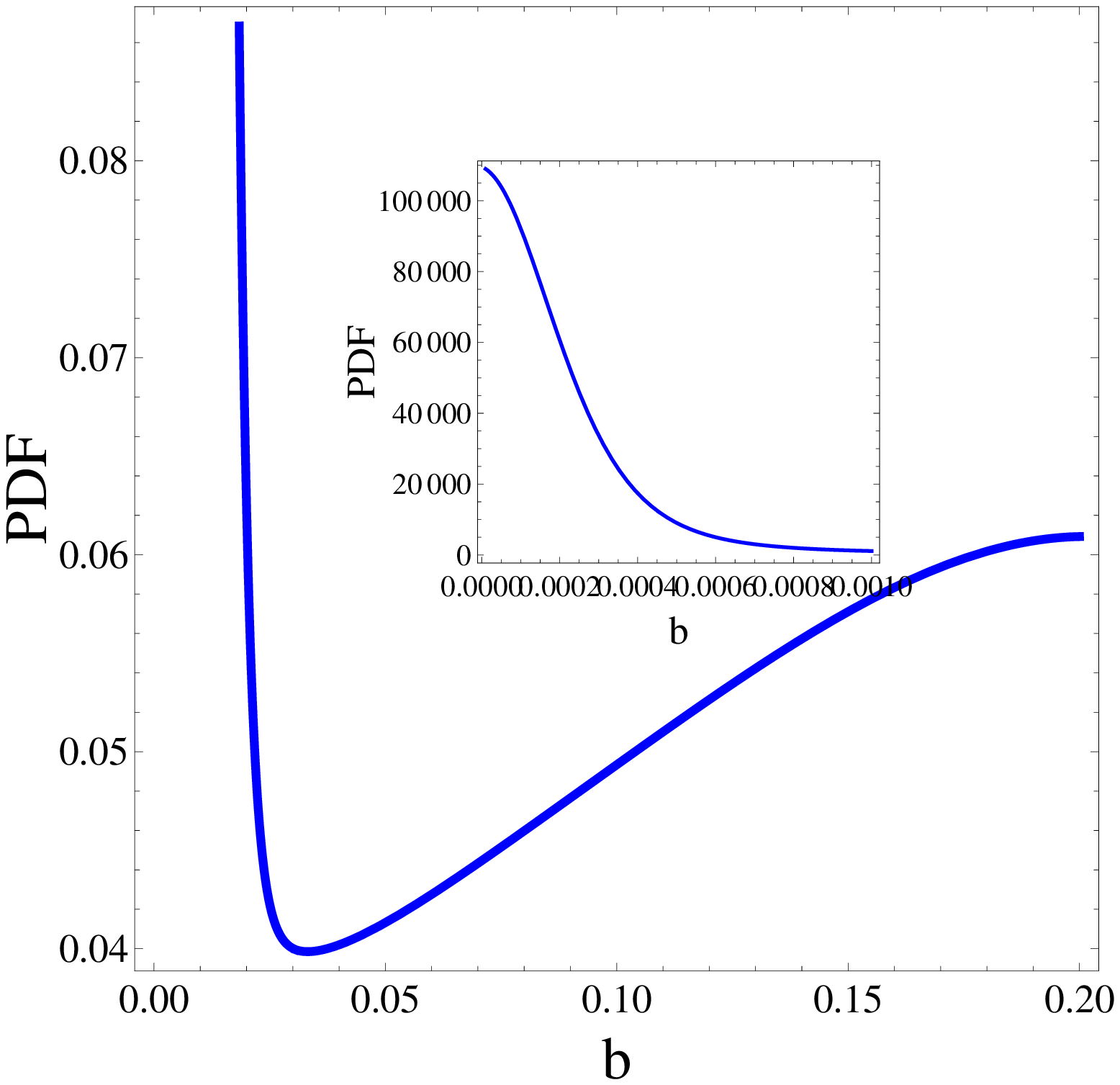}
\end{center}
\caption{The probability density function using 2dFGRS for 
$\Omega_{m0}$(right) and $b$(left). The confidence level 
with 1-$\sigma$, 2-$\sigma$ and 3-$\sigma$.
To build the PDF we marginalize the
free parameters considering the intervals: 
$\Omega_{\rm dm0}$ $\epsilon$ $[0.05, 0.95]$ 
$\Lambda$CDM and $b$ $\epsilon$ 
$[0.001, 0.4]$, 
for each case. As we can see the  model is included,i.e, 
the first figure shows that for a $\Omega_{\rm dm0} \approx$ $0.25$ 
corresponds to a $b \approx 0$.}
\label{fig1-3}
\end{figure}
\begin{table}[!h]
\caption{\label{}
In this table we compile the results of 
observational constraints using different dataset. 
The statistical error is $68.3\%$($1\sigma$) and the results in 
brackets correspond to an error of $2 \sigma$.}
\center
\begin{tabular}{ccccc}
 Data&$\chi^{2}_{min}$ &$\Omega_{\rm dm0}$&$b$ 
\\
\hline
2dFGRS & $21.495$ & $0.0069^{+0.1604 [+0.3400]}_{-0.0069 [-0.0069]}$  & 
$0.001^{+0.002 [+0.003]}_{-0.001 [-0.001]}$
 \\
\hline
SNIa & $542.45$ & $0.21^{+0.02}_{-0.01}$ & $0.48^{+0.31}_{-0.10}$ 
 \\
\hline
R+BAO+H(z)& $8.449$ & $0.28^{+0.07}_{-0.05}$ &  $\approx 0^{+0.006}_{-0.000}$  
\\
\hline
 R+BAO+H(z)+SNIa(Base)& $547.383$ & $0.26^{+0.03}_{-0.05}$ &  
$0.00064^{+ 0.00027}_{-0.00064}$  
\\
\hline
 R+BAO+H(z)+2dfGRS& $24.24$ & $0.28^{+0.07}_{-0.05}$ 
&  $\approx 0^{+0.001}_{-0.000} $  
\\
\hline
R+BAO+H(z)+SNIa+2dfGRS& $563.271$ & $0.26^{+0.05}_{-0.04}$
&  $\approx 0.0001^{+0.003}_{-0.0001}$
\\
\hline
\end{tabular}
\end{table}

\subsection{Combining the Datasets}

We considered that the observational data are independent, 
so we defined the $\chi^{2}_{total}$ as
\beq
\chi^{2}_{total}= \bar{\chi}_{SNIa}^{2} + \chi_{BAO}^{2} + \chi^2_{R} 
+ \chi^{2}_{H}+\chi^{2}_{2dF}
\eeq
The best fit values the model can be determined by minimizing the 
total $\chi^{2}$. For Gaussian distributed measurements, the 
$\chi^{2}$ function is directly  related to the maximum likelihood 
estimator. The likelihood function is determined as
\beq
L(\Omega_{dm0,b})= L_{0}\exp \left(-\frac{\chi^{2}_{total}}{2}\right)\,,
\eeq
where $L_{0}$ is a normalization constant. In order to constraint 
the parameters of our interest, we marginalize over the other parameters.
In the Figs 4, 5 and 6. We displayed the results the $\chi^{2}$ total.
An important issue that needs to be considered is the difference 
between the PDFs of the background tests (BAO+R+H+SNIa) and linear 
perturbations.
In the case of $\Omega_{\rm dm0}$ parameter, we see that the low probability
in the range of $\Omega_{\rm dm0} < 0.1$ eliminates high probability
of this sector in case of perturbations (Fig. 3). Therefore, the result are 
close to $\Lambda$CDM model. Effects of structure formation
may be critical to study the characteristics of models that 
include warm dark matter, where the speed of the particles
is no longer negligible as in the case of $CDM$.
In the table 1 we compile the results of observational constraints at
$1\sigma$ and for the power spectrum, we give also the estimations at $2\sigma$.


\begin{figure}[!h]
\begin{center}
\includegraphics[height= 7. cm,width=7.5cm]{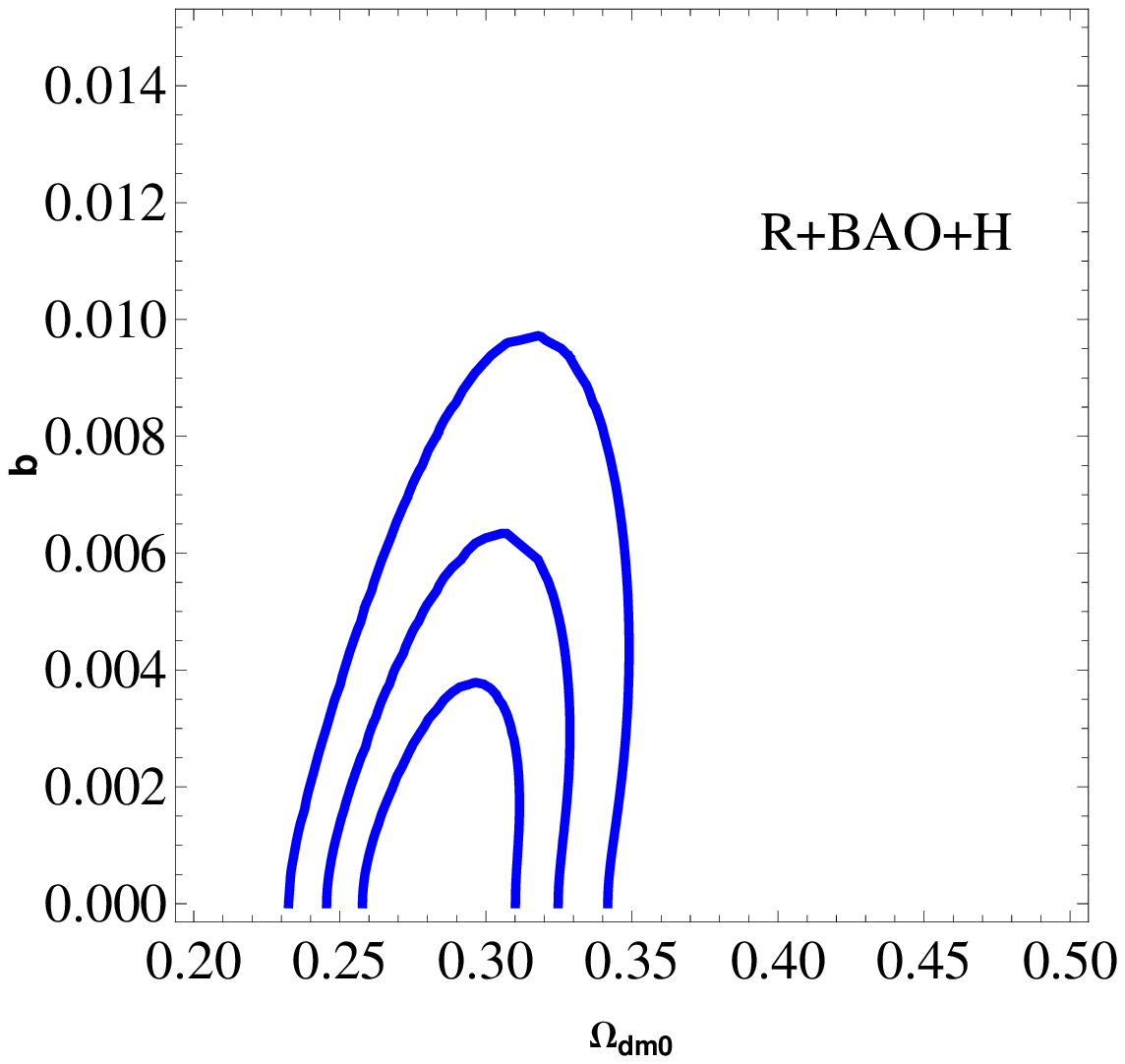}
\includegraphics[height= 7. cm,width=7.5cm]{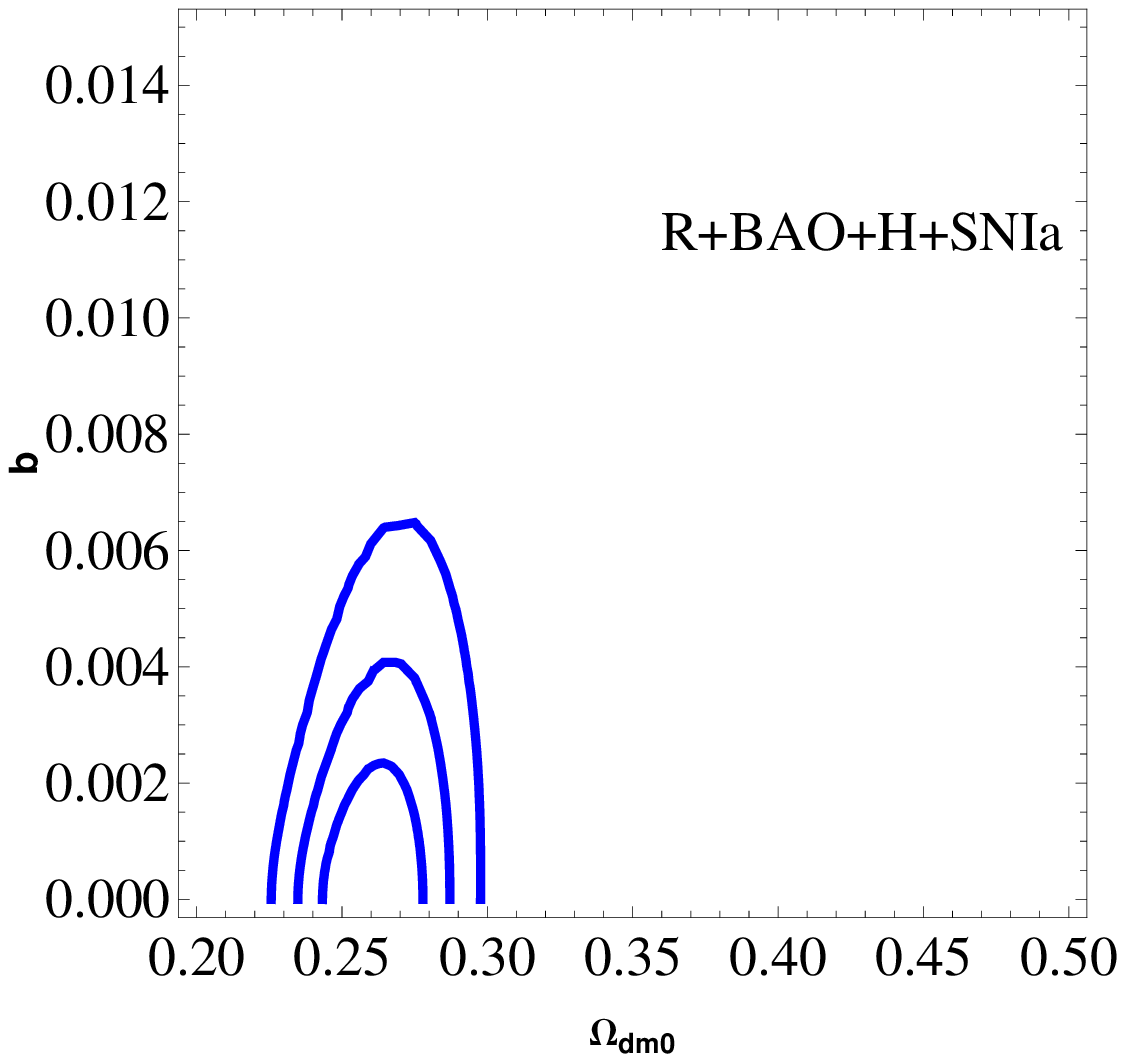}
\includegraphics[height= 7. cm,width=7.5cm]{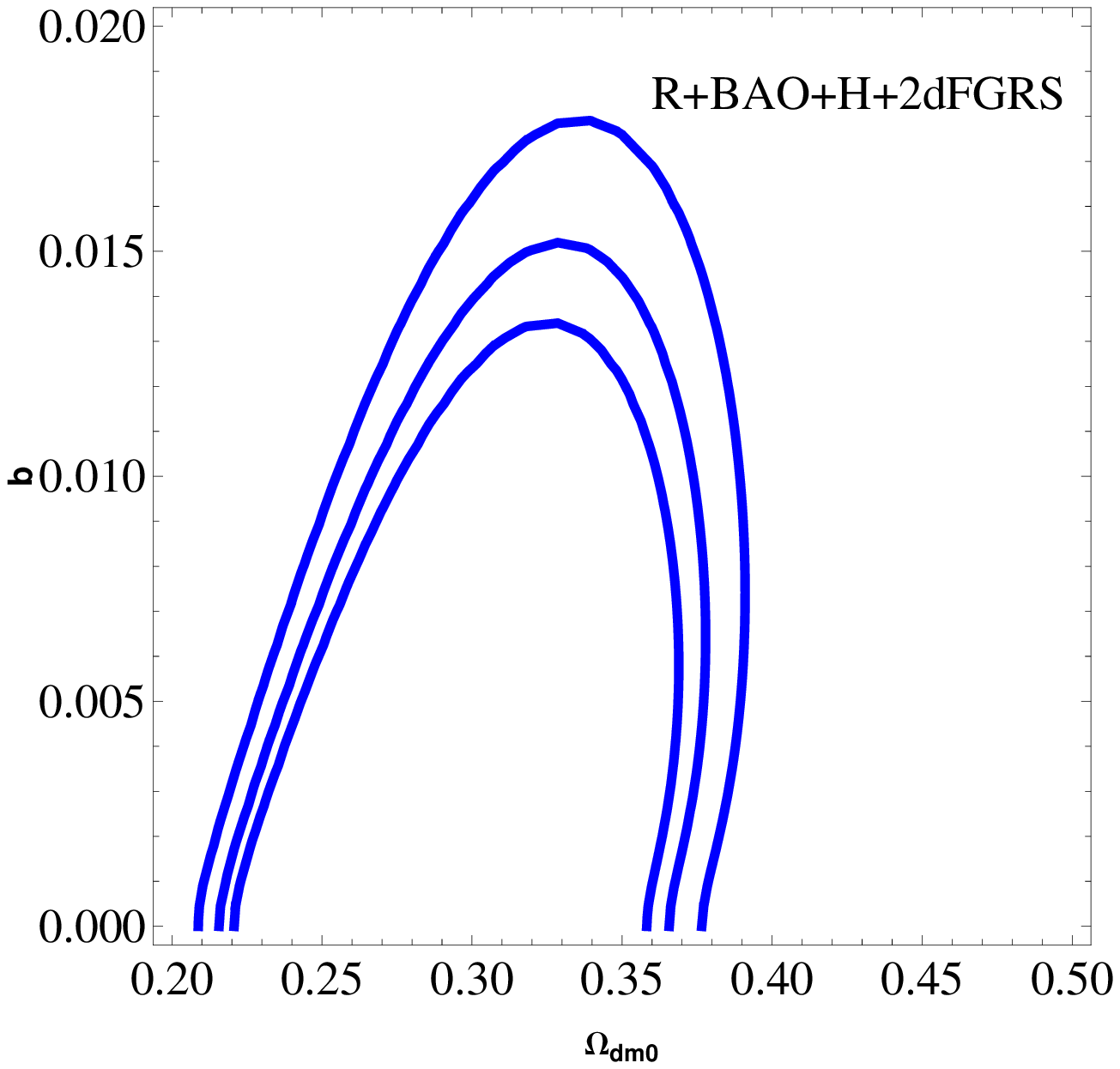}
\includegraphics[height= 7. cm,width=7.5cm]{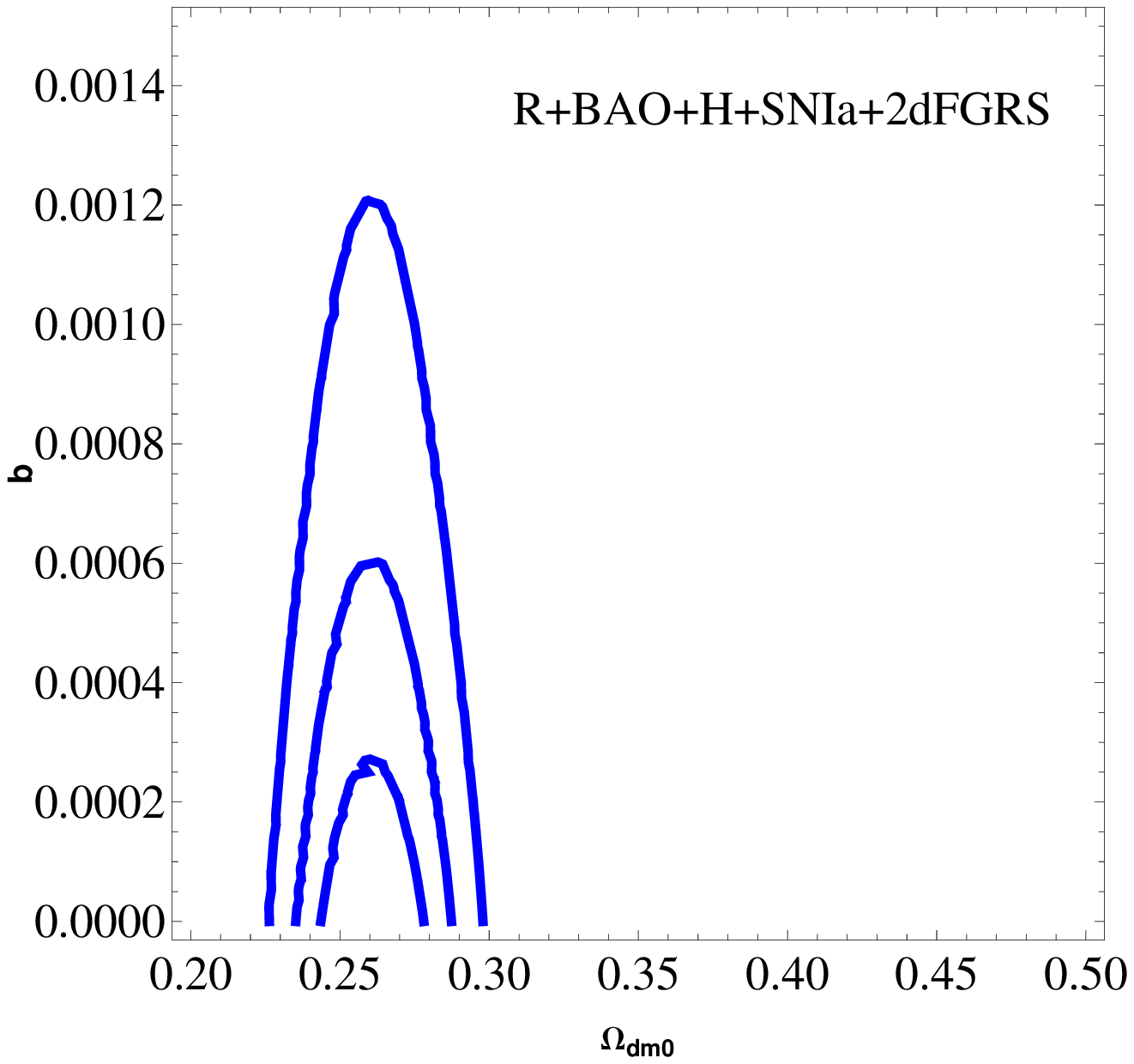}
\end{center}
\caption{Confidence regions at 1-$\sigma$, 2-$\sigma$ and 3-$\sigma$ 
levels from
inner to outer respectively on the ($\Omega_{\rm dm0}$ , $b$) plane for 
our relativistic model in the flat case.
The effect of type Ia supernovae is not apparent 
when we consider only tests of background (H+R+BAO), but it is 
important when combined with power spectrum data, for example,
the constraints on $\Omega_{\rm dm0}$ parameter are more narrow and 
the upper limit for $b$ parameter decreases significantly, an order
of magnitude, compare the last two figures.}
\label{fig1-15}
\end{figure}

\begin{figure}[!h]
\begin{center}
\includegraphics[height= 7.5 cm,width=7.5cm]{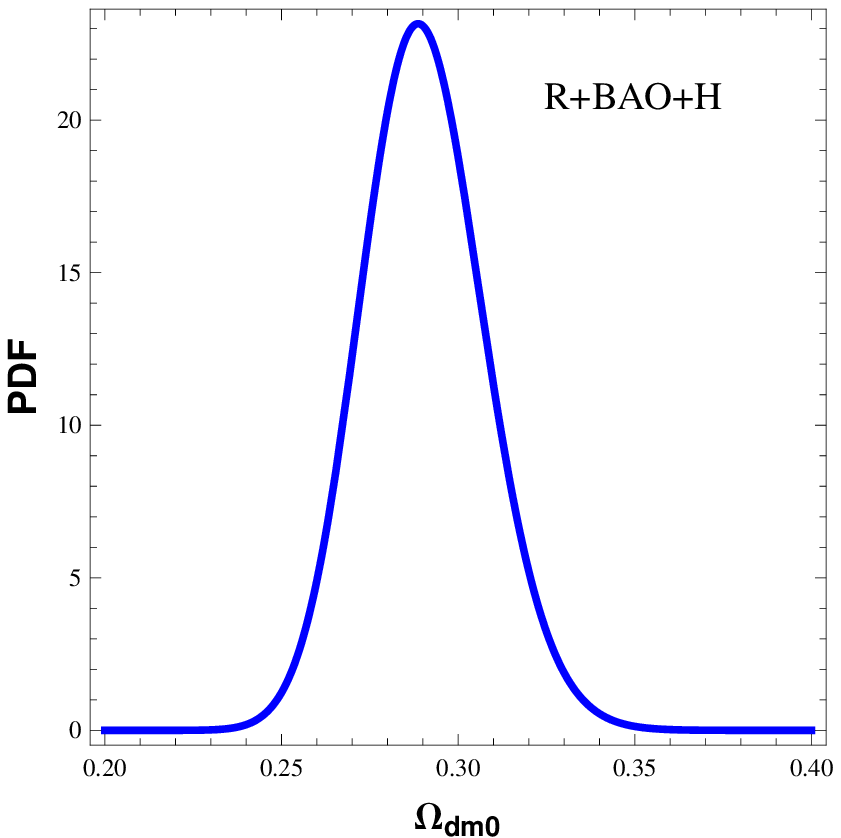}
\includegraphics[height= 7.5 cm,width=7.5cm]{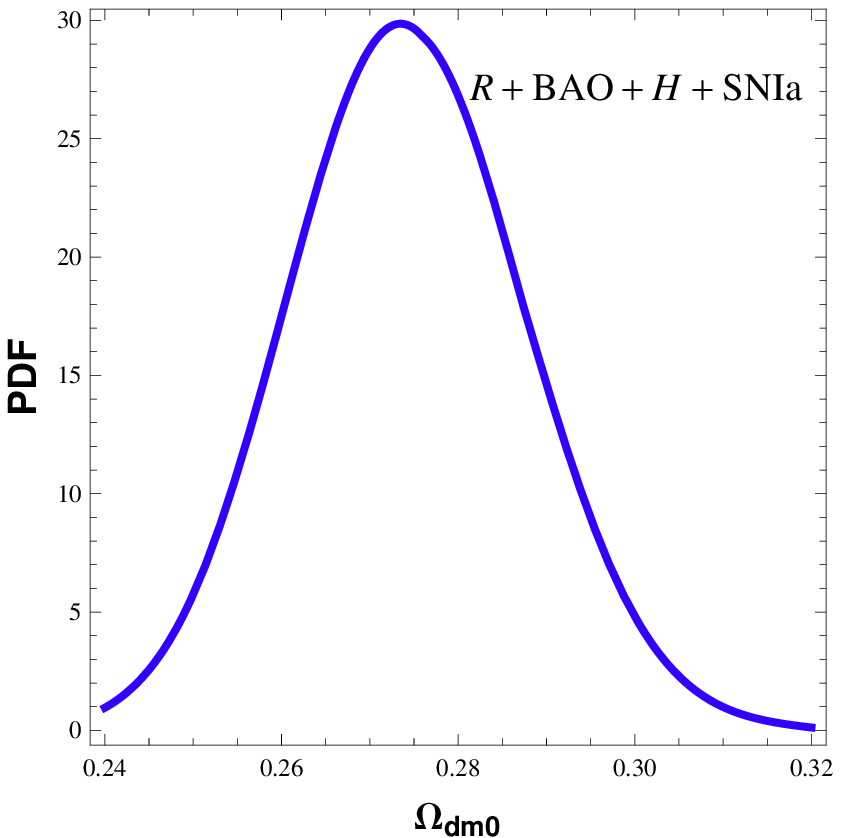}
\includegraphics[height= 7.5 cm,width=7.5cm]{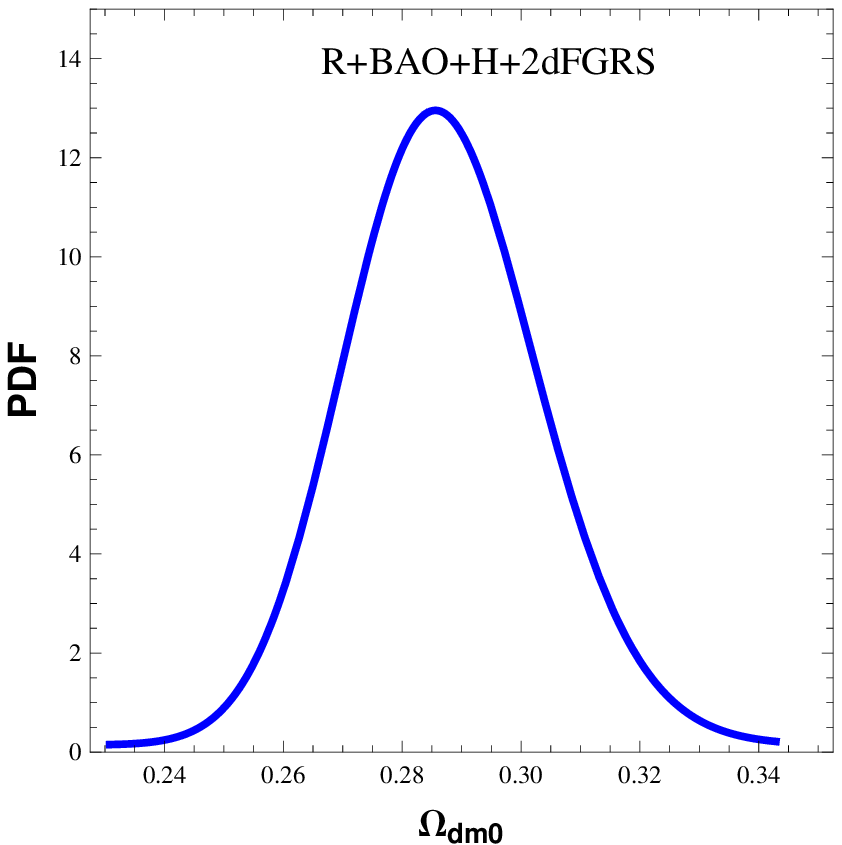}
\includegraphics[height= 7.5 cm,width=7.5cm]{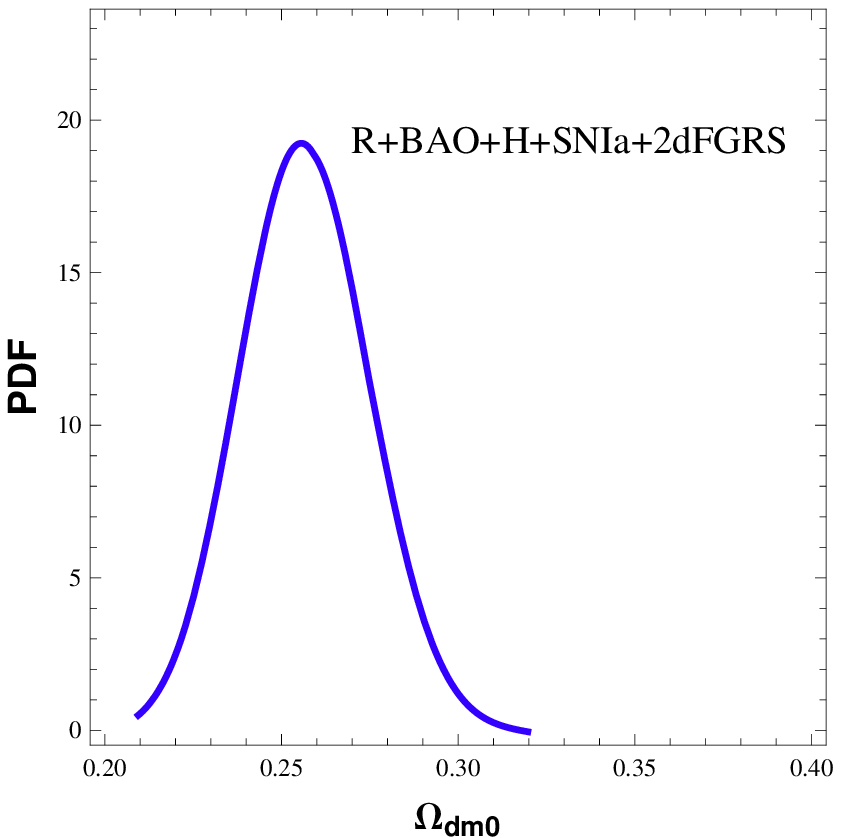}
\end{center}
\caption{The one-dimensional probability distribution function (PDF)
for the $\Omega_{\rm dm0}$ parameter.
The parameter $b$ was marginalized in the
range $[0,1]$.}
\label{fig1-15-N}
\end{figure}

\begin{figure}[!h]
\begin{center}
\includegraphics[height= 7.5 cm,width=6.5cm]{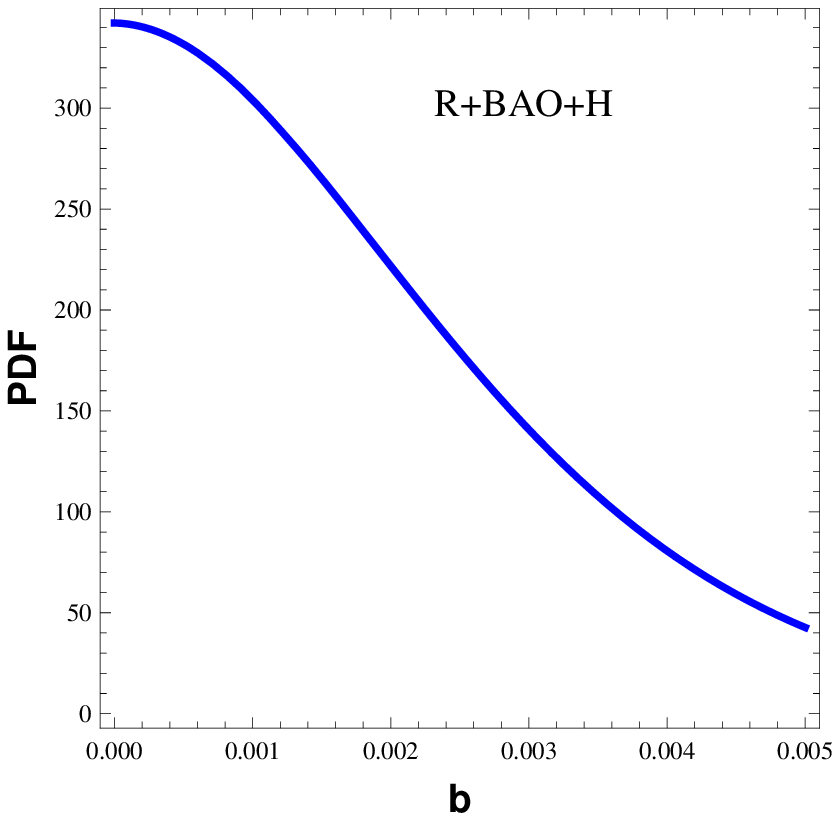}
\includegraphics[height= 7.5 cm,width=6.5cm]{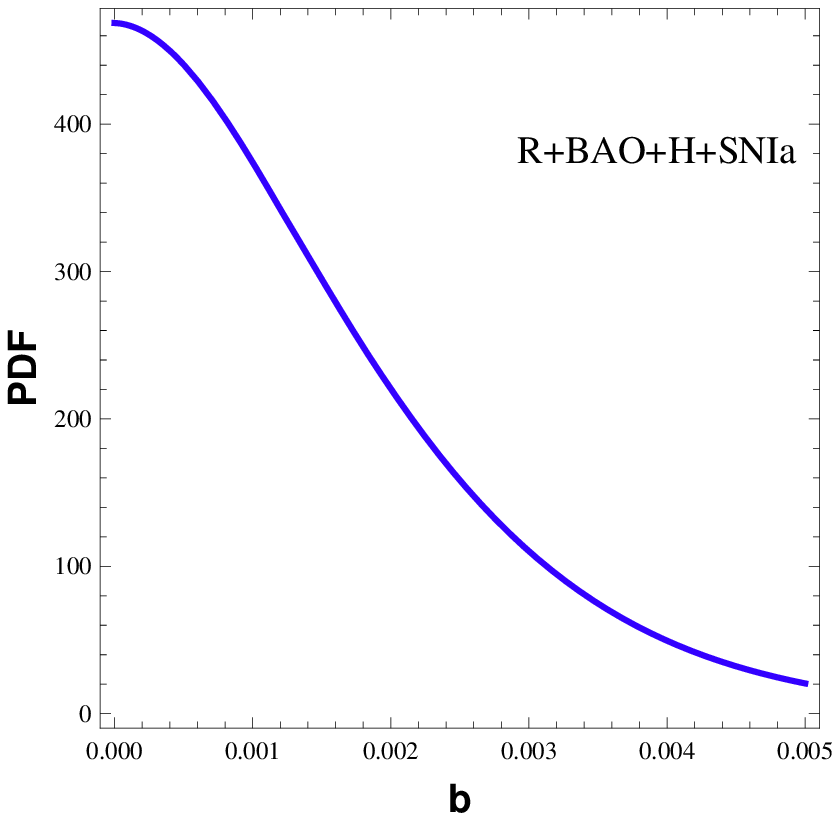}
\includegraphics[height= 7.5 cm,width=6.5cm]{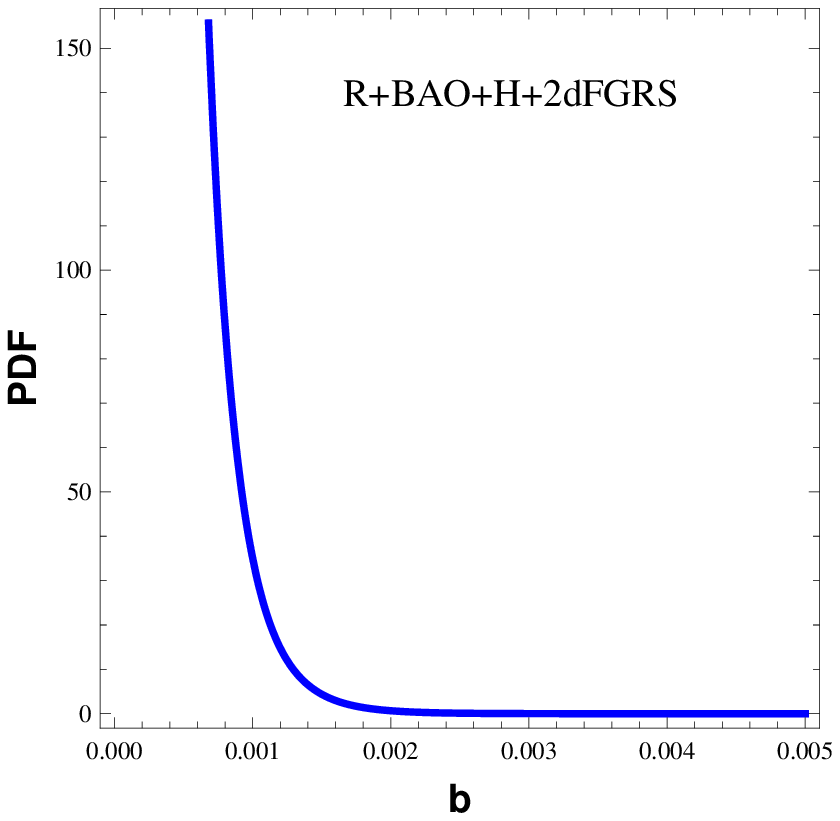}
\includegraphics[height= 7.5 cm,width=6.5cm]{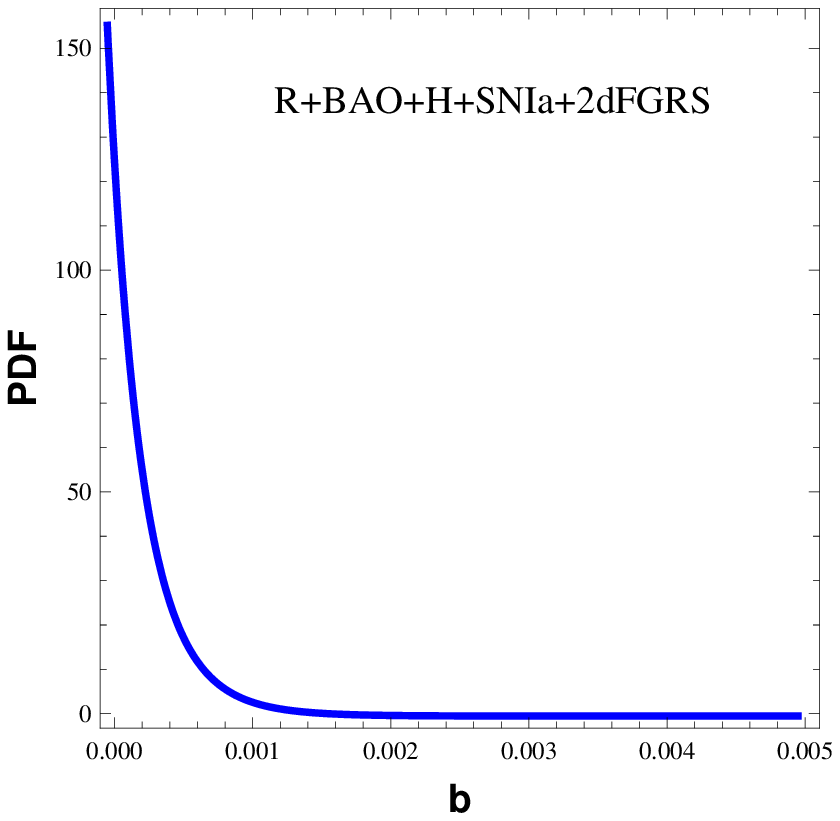}
\end{center}
\caption{The one-dimensional probability distribution function 
(PDF) for the $b$ parameter. The $\Omega_{\rm dm0}$ was marginalized 
in the range \ $[0.05,0.95]$.}
\label{fig1-15-NN}
\end{figure}

\newpage

\section{\bf  Results and Discussion}

The Reduced Relativistic Gas model (RRG) enables one 
to determine the equation 
of state which interpolates between the radiation-dominated 
and the matter-dominated eras. It can represent a warm dark 
matter, characterized by the parameter $b$, where $v \approx bc$. 
The main difference from WDM and CDM come from the behaviour at 
small scales, even during the relativistic phase. WDM is composed 
from lighter particles with respect to the CDM particles, and 
density fluctuations are suppressed at small scales, presenting 
the effect of a free stream. In our case, such small scales effect 
are not relevant, since we are interested in scales corresponding 
to the linear regime, and transition is smooth, not taking into 
account the details of the decoupling of the warm dark matter 
particles. In such a case, the main effect of our model is due 
to the non-zero, even though small, value of the velocity of the 
particles. For our purposes, we can ignore these small scales 
process and take into account only the informations encoded 
in the transfer function for our relevant scales. For very 
large scales, the WDM and CDM transfer functions are essentially 
the same, and small differences appear only at intermediate 
scales, as we could expect. This is in agreement with our results 
which are essentially the same for the WDM and CDM transfer 
function. Hence, the dark component would have a small (but 
non zero) velocity today.

The model has been tested using four 
background observational tests: Supernova type Ia (Union2 sample), 
$H(z)$, CMB ($R$ factor) and BAO. Moreover, a detailed study of 
structure formation at linear level has been performed using the 
2dFGRS data for matter power spectrum. The different tests have 
been crossed in order to obtain a more clear prevision for the 
free parameters, which are essentially the velocity 
parameter for the dark matter particles $b$ and the dark matter
ratio to the critical density $\Omega_{\rm dm0}$. All the analysis 
has been performed using the flat universe prior.

In general, the background tests predict a $\Omega_{\rm dm0}$ very 
similar to the value obtained in the $\Lambda$CDM model, except 
for the CMB $R$ factor which predicts a universe essentially 
dominated by dark matter. For the velocity parameter $b$, the 
background tests predict a maximum probability for a finite 
value at $b \ll 1$, except for BAO, where a large value for $b$ 
is predicted. But the most striking feature of the observational 
tests concerns the formation of structure, for which the maximum
probability for $\Omega_{\rm dm0}$ occurs at a zero value. 
This seems to be a consequence of the restriction of the analysis 
to a linear level, 
since a certain amount of dark matter is necessary in order to 
have the formation of structure process. In any case, the results 
concern a probability distribution: a small amount of dark matter 
is certainly admitted, but much less than that predicted by the 
$\Lambda$CDM model. For the parameter $b$, the PDF is concentrated 
around zero, but there is another region of much smaller, albeit
non-zero, probabilities from $b \sim 0.04$ on. We remember that 
the estimations of warm dark matter in dwarfs spiral galaxies
predict $b \sim 10^{-3}$\cite{vega}. In this context, our model 
seems to be a possible alternative for understanding the formation 
of structures. Additionally as we had seen in the RRG the 
dispersion velocity  
is a free parameter predictable directly and not derived.

The crossing of the background tests with the ones related 
to cosmic perturbations leads to scenarios similar to the 
$\Lambda$CDM model but with a quite large dispersion in the 
estimations for $b$. Here we must distinguish some special 
features. When the SNIa test is not considered, the range of 
allowed value for $b$ becomes larger. The addition of the 
structure formation to the background tests does not change 
considerably the scenario, in spite of the fact that 
test alone would lead to a very small value of $\Omega_{\rm dm0}$. 
This occurs because for small $\Omega_{\rm dm0}$ the PDF is 
essentially zero for the background tests: the crossing of 
perturbative and background tests renders this possibility 
very unlikely. However, considering an error of $2\sigma$ 
(see table 1) our value of $\Omega_{\rm dm0}$ includes the 
value of the $\Lambda$CDM model 
$\Omega_{\rm dm0}=0.229 \pm 0.015$ \cite{komatsu}.

One can resume the main results of this work in the following way. 
The background tests result, without the SN data, is centered in 
the usual values characterizing the $\Lambda$CDM model, that 
means $b \approx 0$ and $\Omega_{\rm dm0} \sim 0.26$, but with a 
non-negligible dispersion. The inclusion of the SN data 
sharpens the confidence region, leading to stronger evidences 
in favor of the $\Lambda$CDM model. In both cases, the inclusion 
of the LSS test enlarges the dispersion mainly in the parameter 
$b$ direction. For the one-dimensional estimations, those 
related to the dark matter density changes very little with, while 
for the $b$ parameter, the inclusion of SN in the background tests 
compress the PDF distribution near $b = 0$, and the inclusion of
the LSS tests accentuates this effect. This seems to be in 
contradiction with the previous two-dimensional description, 
but it must be seen as an effect of the marginalization on the 
density parameter.
\par

In the literature the mechanisms most used for to study warm 
dark matter has been based in a distribution function of the form 
$f_{x}(v)=\frac{\be}{\ep^{p/\al T_{\ga}}+1}$ \ \cite{colombi} 
\ where 
$T_{\ga}$ is the photon temperature, $v=p/(p^{2}+m_{x}^{2})^{1/2}$, 
and three parameters $\alpha$, $\beta$, and $m_{x}$. The power
spectrum today and the relation between pressure and energy density 
is given with a function of this distribution. In this context the 
transfer function can be approximated by the fitting function 
$T(k)=[1+(\alpha k)^{2.24}]^{-4.46}$ \cite{bode} where $\alpha$ 
is a function of free parameters of WDM. We have used this 
function with $m_{x}=1keV$ and $m_{x}=0.25keV$ 
and observe that the obtained results are very close to 
those obtained with the BBKS function, therefore we our results
are not affected significantly by the scale of mass.

It is important to remember that the standard model, 
$\Lambda$CDM, appears to be in conflict with observations on 
subgalactic scales. There are two major conflicts between 
$\Lambda$CDM models and observations in the Local Universe.
First, the inner mass density profiles of simulated dark matter 
halos are more cuspy than inferred from the rotation curves of 
dwarfs and low surface brightness galaxies. Currently, this 
conflict seems well established \cite{moore}. Therefore, the 
$\Lambda$CDM models in its current form cannot provide the 
unknown mass that surrounds galaxies. The second problem is 
related with the N-body simulations of $\Lambda$CDM models 
that predict large numbers of low  mass halos greatly in excess 
when compared the observed number of satellite galaxies in the 
Local group. These and other issues suggest that it is necessary 
to investigate different theoretical possibilities. In this paper 
we have shown that, in part of linear perturbations alone, our 
$RRG$ model for a WDM provides results which are different 
from the standard $\La$CDM model. It is possible, in principle,
that the parameter estimations from other observational data 
can be modified by the quantum corrections in the form discussed 
in \cite{Gruni} and \cite{CCG}, such that eventually 
we may converge to an alternative cosmic concordance model. 
The present investigation can be seen as a first step towards 
this program.  

It is clear that to better quantify these qualitative 
expectations, one needs to take into account the nonlinear 
effects, including the dynamics of the collapse and the 
halo concentrations, e.g., by using algorithm suggested 
in Ref. \cite{eke}, etc. Another important thing to do is 
to consider the other sets of observational tests, like the 
integrated Sachs-Wolfe effect, in the first order in quantum
corrections and see whether they are compatible with the 
small $\Omega_{0dm}$ and small WDM model.

\section*{\large\bf Acknowledgments.}
The work of J.F. has been supported in part by CNPq
and by FAPES. The work of A.M.V.T. and I.Sh. has been supported 
in part by CNPq and FAPEMIG, I.Sh. also was supported by ICTP
Senior Associated Membership Program. 

\newpage
\section*{\large\bf  Appendix}

In this appendix, we display the one dimensional PDFs and the 
bi-dimensional confidence levels for each background test taken 
separately.

\begin{figure}[!h]
\begin{center}
\includegraphics[height= 4.7 cm,width=4.5cm]{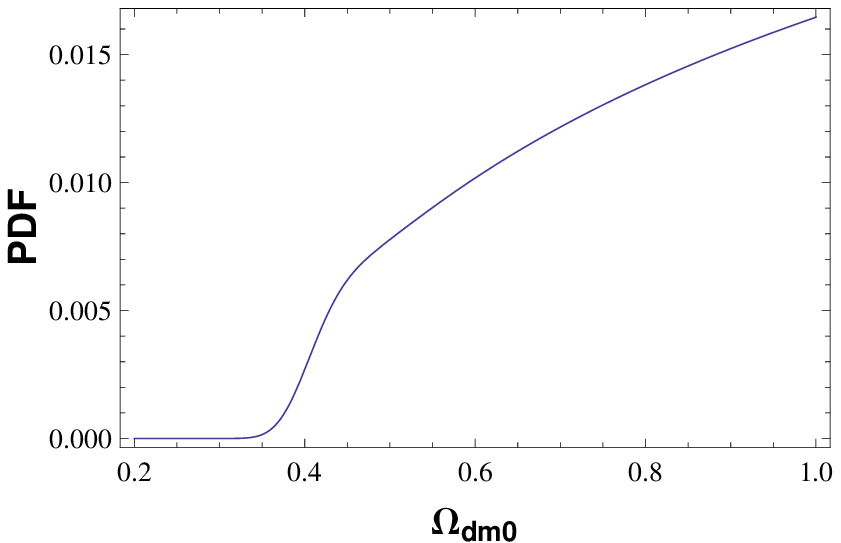}
\includegraphics[height= 4.69 cm,width=4.5cm]{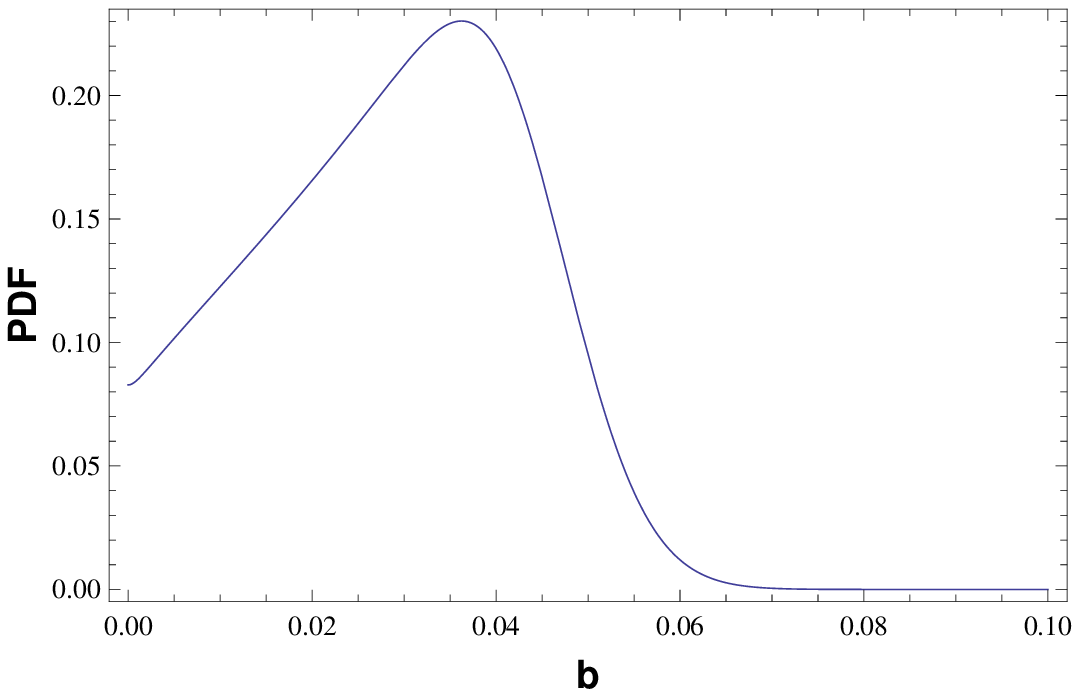}
\includegraphics[height= 4.7 cm,width=4.5cm]{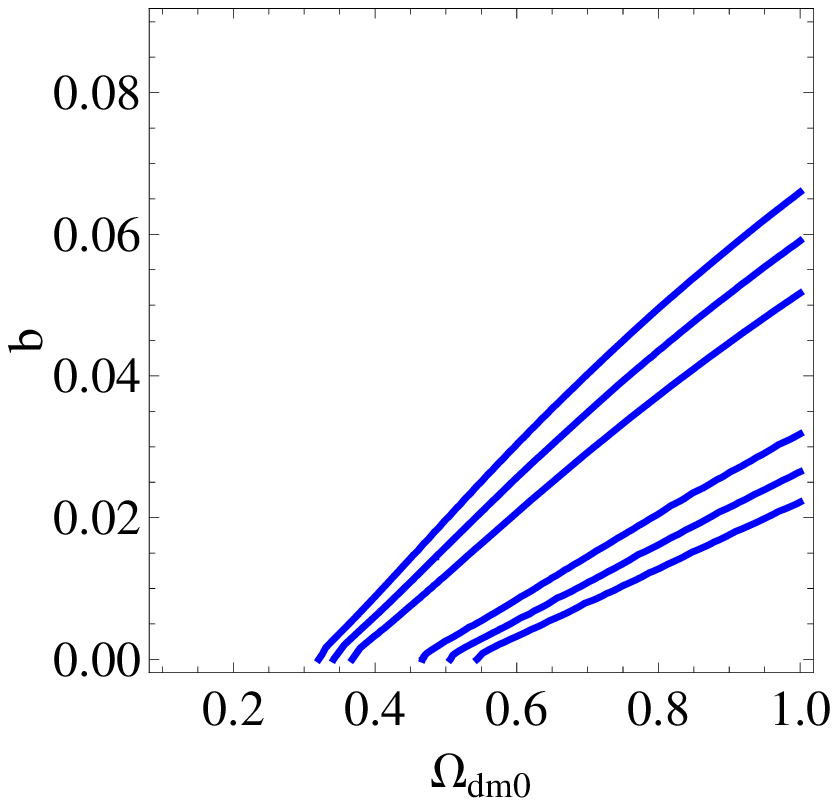}
\end{center}
\caption{The PDFs using CMB. The confidence level 
with 1-$\sigma$, 2-$\sigma$ and 3-$\sigma$(left)}
\label{fig1}
\end{figure}

\begin{figure}[htb]
\begin{center}
\includegraphics[height= 5.0 cm,width=4.5cm]{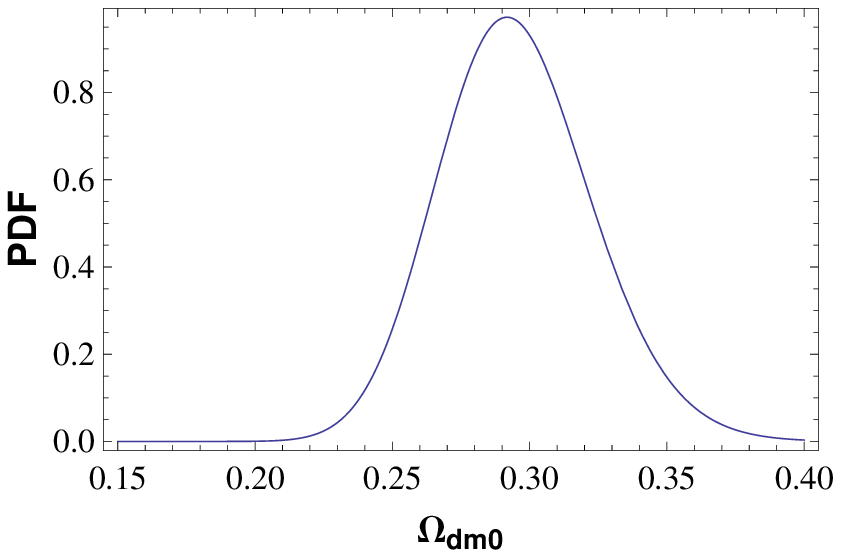}
\includegraphics[height= 5.0 cm,width=4.5cm]{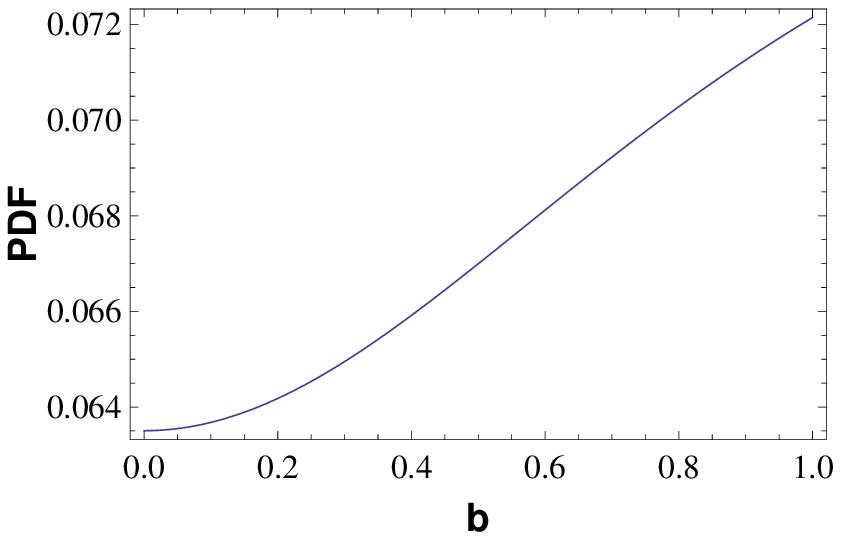}
\includegraphics[height= 5.0 cm,width=4.5cm]{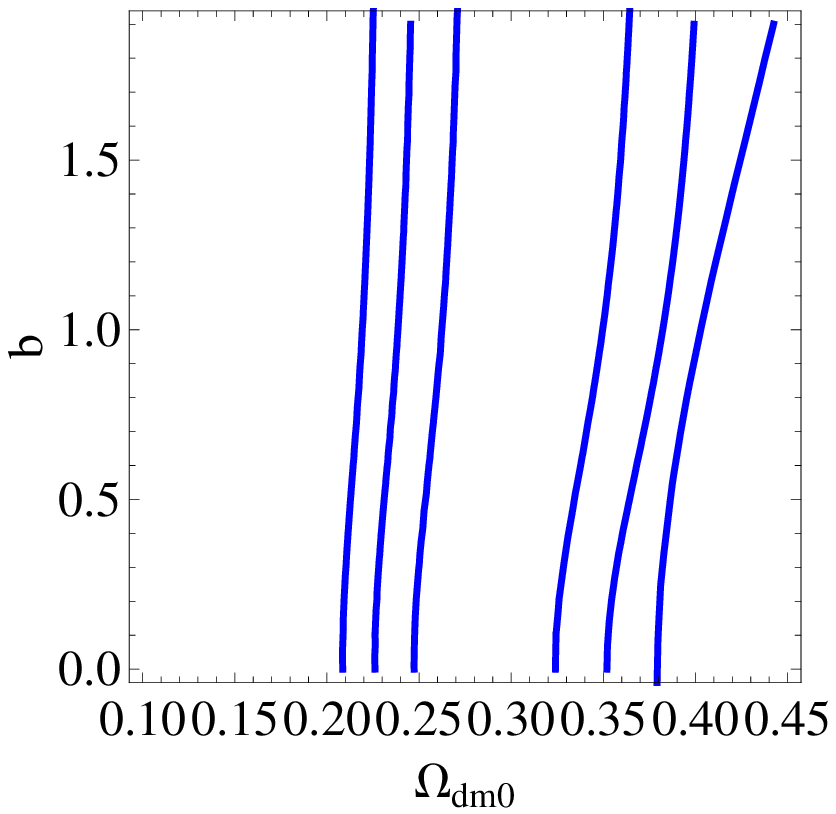}
\end{center}
\caption{The PDFs using BAO. The confidence level 
with 1-$\sigma$, 2-$\sigma$ and 3-$\sigma$(left)}
\label{fig1-1-N}
\end{figure}

\newpage

\begin{figure}[htb]
\begin{center}
\includegraphics[height= 4.45 cm,width=4.5cm]{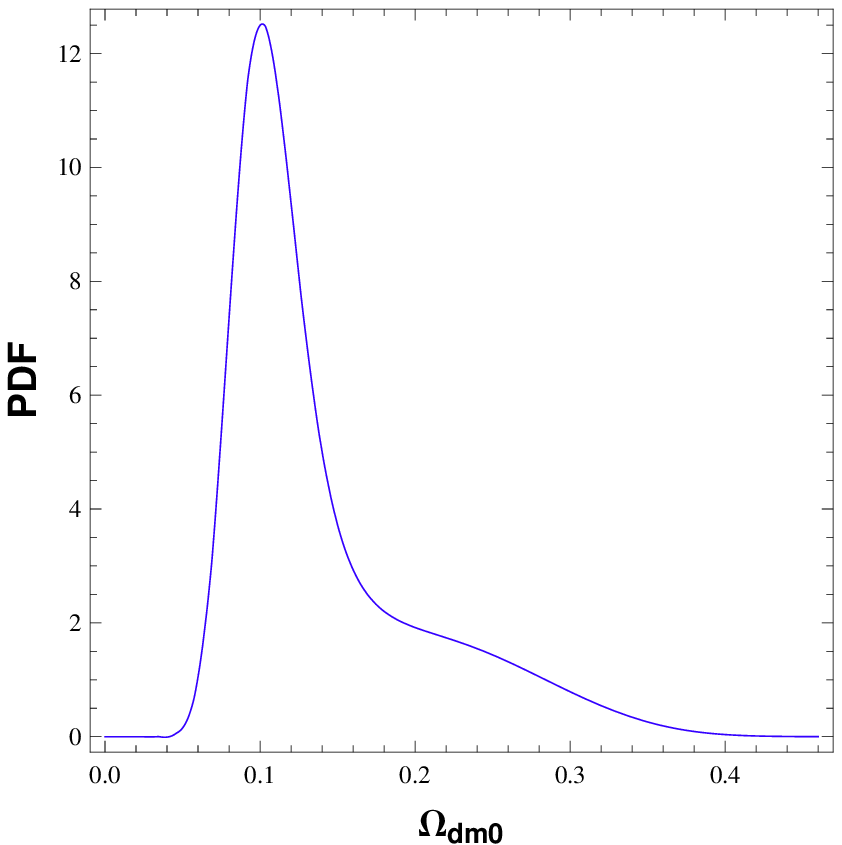}
\includegraphics[height= 4.4 cm,width=4.5cm]{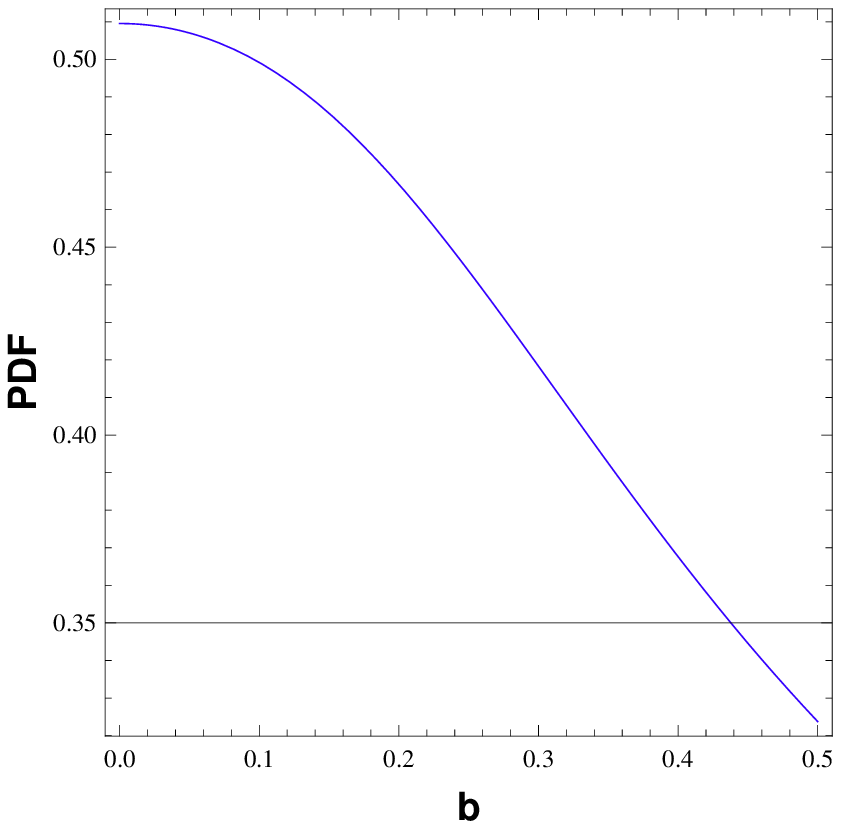}
\includegraphics[height= 4.42 cm,width=4.5cm]{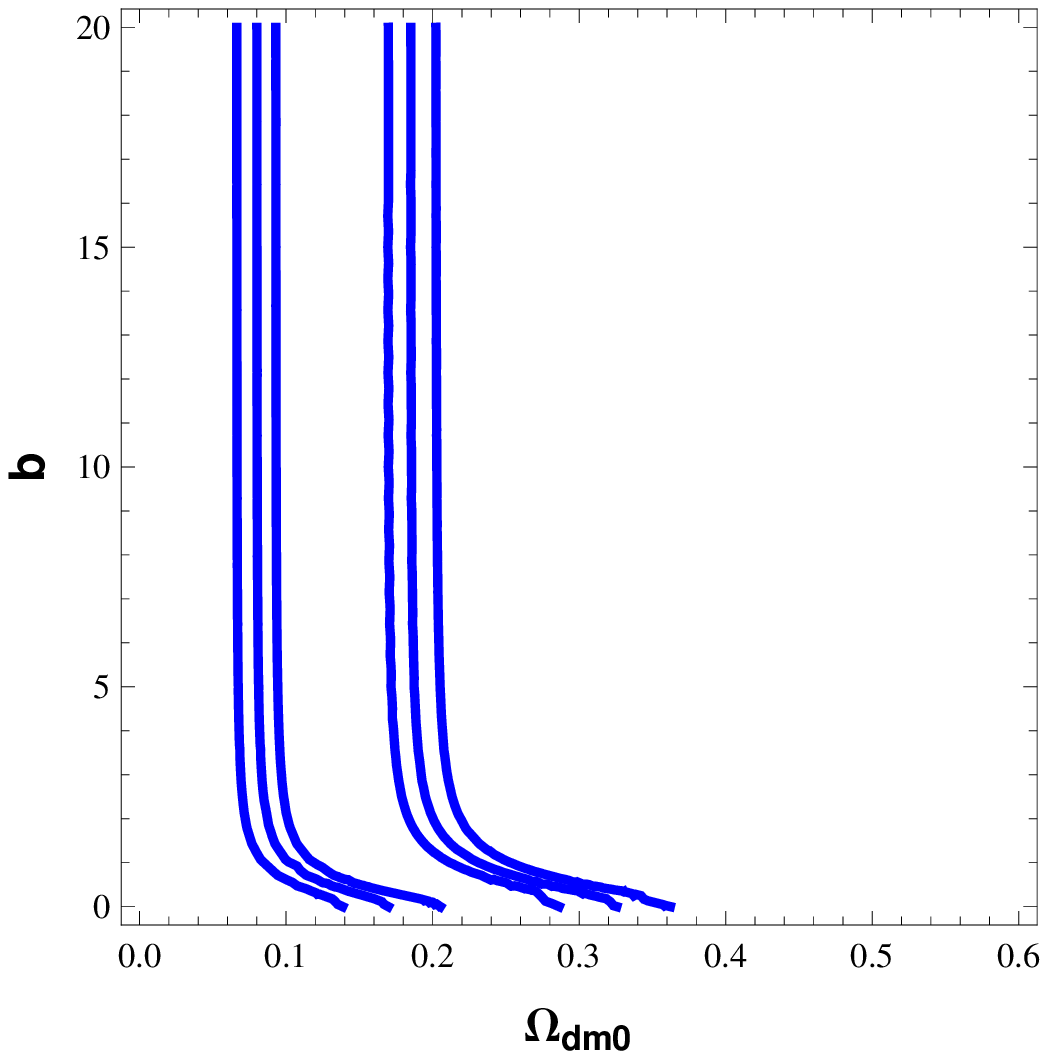}
\end{center}
\caption{The PDFs using H(z) data. The confidence level 
with 1-$\sigma$, 2-$\sigma$ and 3-$\sigma$(left)}
\label{fig1-2}
\end{figure}

\begin{figure}[htb]
\begin{center}
\includegraphics[height= 4.5 cm,width=4.5cm]{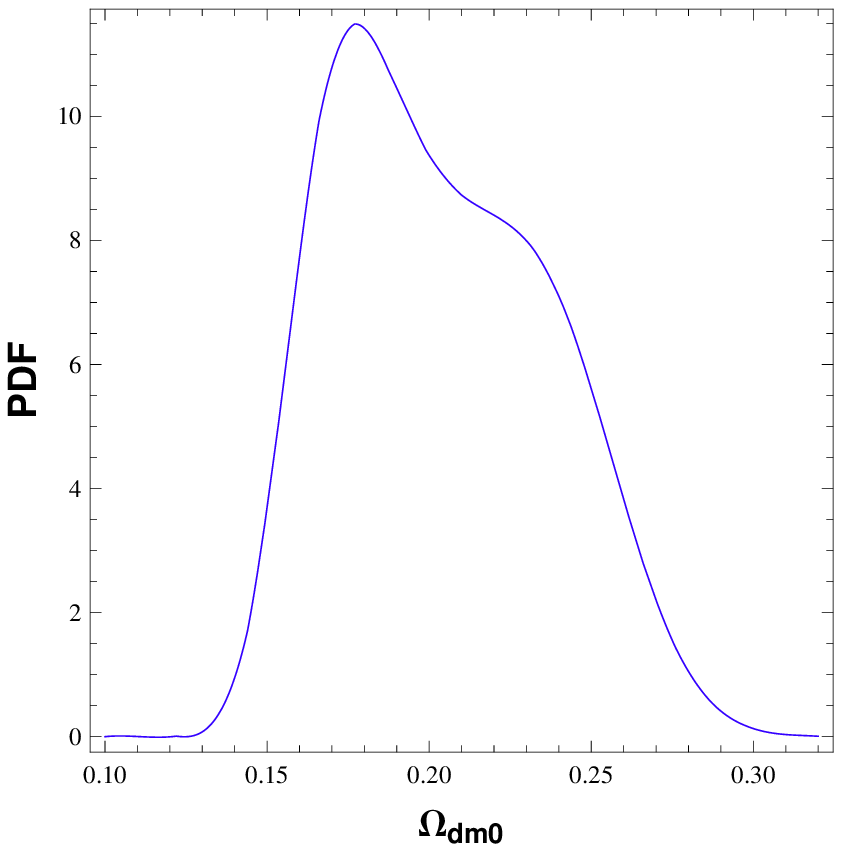}
\includegraphics[height= 4.55 cm,width=4.5cm]{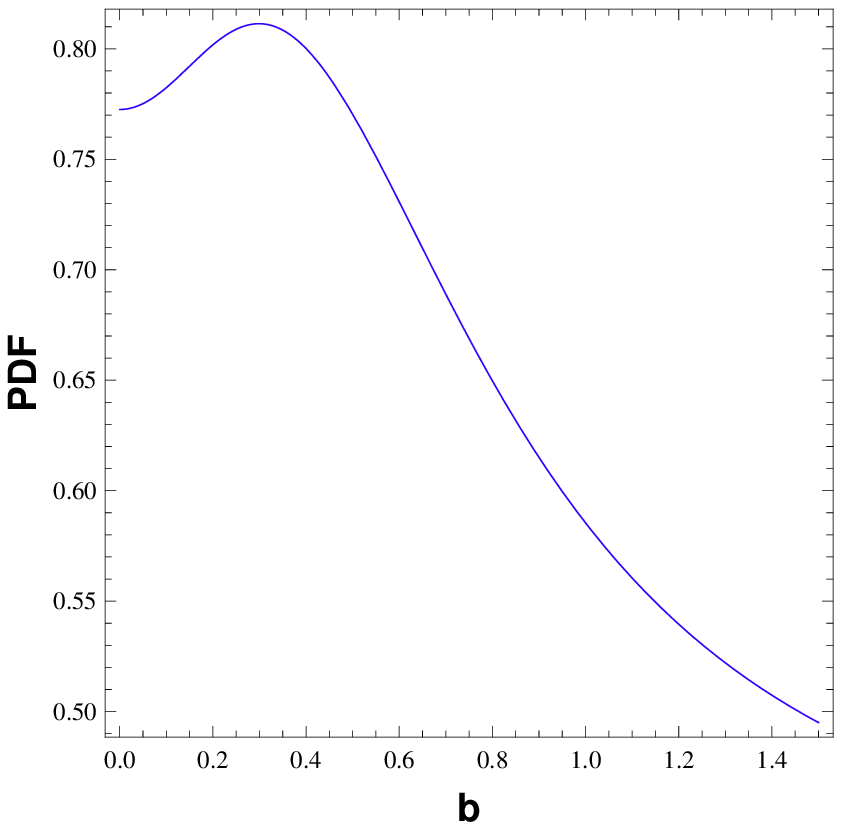}
\includegraphics[height= 4.60 cm,width=4.5cm]{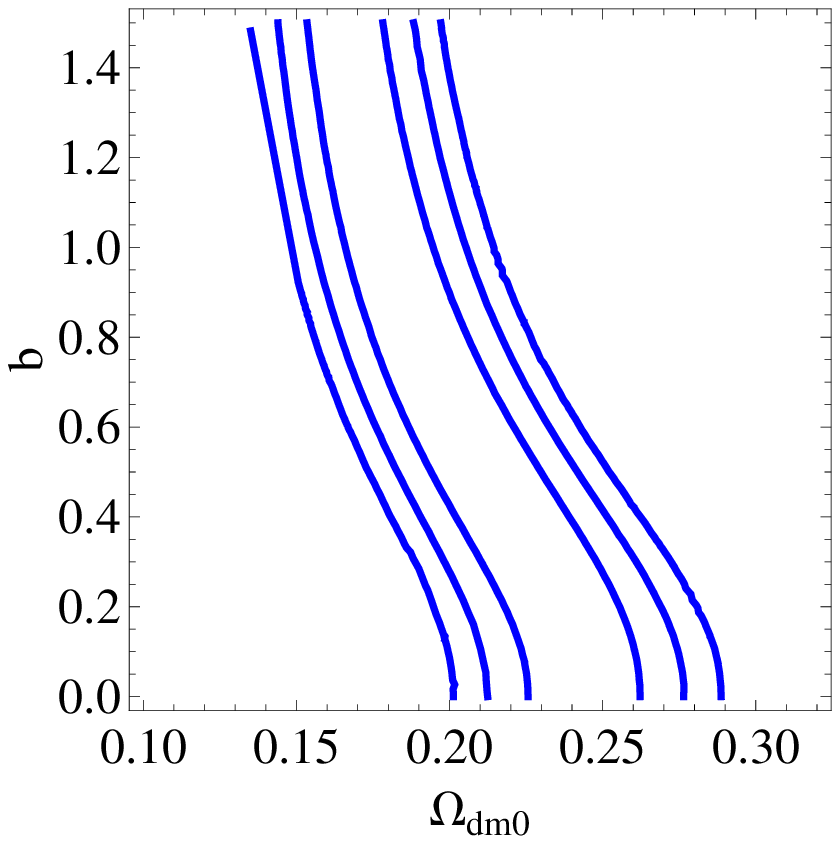}
\end{center}
\caption{The PDFs using SNIa Union 2. The confidence level 
with 1-$\sigma$, 2-$\sigma$ and 3-$\sigma$(left)}
\label{fig1-2-N}
\end{figure}
\newpage



\begin{thebibliography}{10}

\bibitem{riess}
A.G. Riess {\it et al.} Astron. J. {\bf 116}, 1009(1998), e-Print: 
astro-ph/9805201; Astrophys. J. {\bf 699}, 539(2009), 
e-print: arXiv:0905.0695; 
S. Perlmutter, Astrophys.J. {\bf 517}, 565(1999), e-print: astro-ph/9812133.

\bibitem{hannestad} S. Hannestad, Int. J. Mod. Phys. {\bf A21}, 1938(2006).

\bibitem{weinberg89} S. Weinberg, Rev. Mod. Phys. {\bf 61}, 1(1989).

\bibitem{nova} I.L. Shapiro, J. Sol\`{a},
JHEP {\bf 02}, 006(2002).

\bibitem{KT} E. Kolb and M. Turner, {\sl The Very Early Universe}, 
Addison-Wesley, New York(1994); J. Binney and 
S Tremaine, { \sl Galactic Dynamics}, Princeton UP(1988).

\bibitem{Dod} S. Dodelson, {\sl Modern Cosmology}, 
Academic Press, New York(2003).

\bibitem{Mukh}
M. Weber, W. de Boer, Astron. Astrophys. { \bf509}, A25 (2010) 

\bibitem{CoLu}
J. Angle et al. Phys.Rev.Lett. {\bf100}, 021303 (2008)

\bibitem{DM1} J.A. Tyson, G. P. Kochanski, I. P. Dell'Antonio, 
Astrophys.J. {\bf498}, L107 (1998); 

\bibitem{DM2}
A. Pinzke, C. Pfrommer, L. Bergstrom. [arXiv:1105.3240]

\bibitem{susyDM}
J. L. Tinker, Astrophys.J. {\bf688}, 709 (2008) 

\bibitem{Roos} 
I.A. Yegorova, A. Pizzella, P. Salucci. 
[arXiv:1106.5105]

\bibitem{first} 
S. Dodelson and L.M. Widrow, Phys. Rev. Lett. {\bf 72}, 17(1994).

\bibitem{dodelson} S. Colombi, S. Dodelson and L.M. Widrow,
Astrophys. J. {\bf 458}, 1(1996).

\bibitem{FlaFlu} 
G. de Berredo-Peixoto, I. L. Shapiro and F. Sobreira,
Mod. Phys. Lett. {\bf 20A}, 2723(2005).

\bibitem{sWIMPs}
J.C. Fabris, I.L. Shapiro and F. Sobreira,
JCAP {\bf 02}, 001(2009), e-print: arXiv:0806.1969.

\bibitem{Sakharov} A.D.Sakharov, 
Soviet Physics JETP, 22, 241 (1966) 
[Russian original: ZhETF, 49, 345 (1965)]; 

\bibitem{Grishchuk} L.P. Grishchuk, 
{\it Cosmological Sakharov Oscillations and Quantum Mechanics 
of the Early Universe}, arXive:1106.5205; \
Talk at the Special Session of the Physical Sciences Division 
of the Russian Academy of Sciences, Moscow, 25 May 2011. 

\bibitem{szalay} J.R. Bond and A.S. Szalay, Astrophys. J. 
{\bf 274} 443(1986).

\bibitem{ostriker1} Z. Haiman, R. Barkana and J.P.  Ostriker,
{\it Warm Dark Matter, Small Scale Crisis, and the High Redshift 
Universe}, e-print: astro-ph/0103050.

\bibitem{ostriker2} P. Bode, J.P. Ostriker and N. Turok, 
Astrophys. J. {\bf 556}, 93(2001).

\bibitem{ostriker3} R. Barkana, Z. Haiman and J. P. Ostriker,
e-print: astro-ph/0102304.

\bibitem{RotCurves} D.C. Rodrigues, P.S. Letelier and I.L. Shapiro, 
JCAP {\bf 04}, 020(2010); e-print: arXiv: 0911.4967. 

\bibitem{Gruni}
I.L. Shapiro, J. Sol\`{a}, H. \v{S}tefan\v{c}i\'{c},
JCAP {\bf 0501}, 012(2005).

\bibitem{CCG}
J. Grande, J. Sol\`{a}, J.C. Fabris and I.L. Shapiro,
Class. Quant. Grav. {\bf 27}, 105004(2010).

\bibitem{CCwave} J.C. Fabris, I.L. Shapiro and J. Sol\`{a}, 
JCAP {\bf 0702} 016(2007); e-print: gr-qc/0609017. 

\bibitem{LRL} C. Farina, W.J.M. Kort-Kamp, S. Mauro
and I.L. Shapiro, 
Phys. Rev. {\bf D83} 124037(2011).

\bibitem{bbks} J.M. Bardeen, J.R. Bond, N. Kaiser, A.S. Szalay,
Astrophys. J. {\bf 304} 15(1986). 

\bibitem{bode}
P. Bode, J. P. Ostriker and N. Turok, Astrophys. J. {\bf 556}, 93 (2001).
\bibitem{Riotto-Lyman}
M. Viel, J. Lesgourgues, M.G. Haehnelt, S. Matarrese and A. Riotto, 
Phys. Rev. {\bf D71}, 063534(2005), e-print: astro-ph/0501562; \ \ 
%
Phys. Rev. Lett. {\bf 97}, 071301(2006), e-print: astro-ph/0605706. 

\bibitem{komatsu}
E. Komatsu et al;  
Astrophys.J.Suppl. {\bf 192} 18(2011) e-print: arXiv:1001.4538.

\bibitem{pn}
S. Nesseris, L. Perivolaropoulos, JCAP {\bf 01}, 018(2007). 

\bibitem{eisenstein}
D.L. Eisenstein et al. (SDSS) 
ApJ {\bf 633}, 560(2005); \ E-print: astro-ph/0501171.

\bibitem{ji} R. Jimenez, L. Verde, T. Treu and D. Stern, 
Astrophys. J, {\bf 593}, 622(2003).

\bibitem{si}
J. Simon, L. Verde, R. Jimenez, Phys. Rev. D 71, 123001(2005).

\bibitem{stern}
D. Stern, R. Jimenez, L. Verde, M. Kamionkowski and S.A. Stanford, 
JCAP {\bf 2}, 8(2010), e-print: arXiv:0907.3149.

\bibitem{gazt}
E. Gaztanaga, A. Cabre and L. Hui, 
Mon.\ Not.\ Roy.\ Astron.\ Soc.\ {\bf 399}, 1663(2009),
e-print: arXiv:0807.3551.

\bibitem{amanullah} R. Amanullah {\it et al.}  
Astrophys. J, {\bf 716}, 712(2010).

\bibitem{kessler} 
R. Kessler {\it et al.} Astrophys. J. {\bf 185}, 32(2009).

\bibitem{kowalski} 
M. Kowalski {\it et al.} Astrophys. J. {\bf 686}, 749(2008). 

\bibitem{cole} S. Cole et al.; 
Mon.\ Not.\ Roy.\ Astron.\ Soc.\  {\bf 362}, 505 (2005); 
Astrophys.\ J.\  {\bf 549}, 669(2001), e-print: astro-ph/0006436; 
W.~J.~Percival {\it et al.}  [The 2dFGRS Team 
  Collaboration], Mon.\ Not.\ Roy.\ Astron.\ Soc.\  {\bf 337}, 1068(2002),
 e-print: astro-ph/0206256.

\bibitem{vega}
H. J. de Vega and N. G. Sanchez, 
Int. J. Mod. Phys.A {\bf 26}, 1057(2011);

H. J. de Vega, P. Salucci and N. G. Sanchez, e-print: arXiv:1004.1908.

\bibitem{colombi}
S. Colombi, S. Dodelson and L. M. Widrow, Astrophys. J. {\bf  458}, 1(1996).

\bibitem{moore}
R. K. de Naray, T. Kaufmann, MNRAS {\bf414}, 3617 (2011); 
A. Del Popolo, MNRAS {\bf408}, 1808 (2010); 
F. Donato, et al. MNRAS {\bf397}, 1169 (2009);
G. Ogiya and M. Mori, e-Print: arXiv:1106.2864

\bibitem{eke}
V.R. Eke, J.F. Navarro and M. Steinmetz, Astrophys. J. { \bf 554}, 114(2001).

\vskip 10mm



\end{thebibliography}
\end{document}